\documentclass[preprint,10pt]{aastex}
\usepackage{emulateapj5}
\newcommand{\degree}{\mbox{$^{\circ}$}}

\newcommand{\as}{\mbox{\arcsec}}

\newcommand\cmv{\mbox{cm$^{-3}$}}

\def\lsim {$\rlap{\raise.4ex\hbox{$<$}}\lower.55ex\hbox{$\sim$}\,$}

\newcommand{\lsun}{\mbox{L$_\odot$}}% Lsun
\newcommand{\msun}{\mbox{M$_\odot$}}% Msun
\newcommand{\ta}{{$T_A^*$}}

\newcommand{\tr}{\mbox{$T_R^*$}}

\newcommand{\lbol}{\mbox{$L_{bol}$}} % bolometric luminosity
\newcommand{\tbol}{\mbox{$T_{bol}$}} % bolometric temperature

\newcommand{\mean}[1]{\mbox{$\langle#1\rangle$}} %generic mean for defined qu.
\newcommand{\av}{\mbox{$A_V$}} % Visual Extinction
\newcommand{\lbolsmm}{\mbox{$\frac{L_{bol}}{L_{smm}}$}} % Ratio of Lbol to Lsmm
\newcommand{\form}{H$_2$CO}

\newcommand{\ammonia}{\mbox{{\rm NH}$_3$}}

\newcommand{\hcop}{HCO$^+$}

\newcommand{\nthp}{N$_2$H$^+$}
\newcommand{\cts}{C$_3$S}
 
\input{epsf}

\newcommand{\Snu}{\mbox{$S_{\nu}$}}

\newcommand{\kappanu}{\mbox{$\kappa_{\nu}$}}

\newcommand{\sisrf}{\mbox{s$_{\rm{isrf}}$}}

\begin{document}

%%%%%%%%%%%%%%%%%% title %%%%%%%%%%%%%%%%%%%%%%%%%%%%%%%%%%%%%%%%
  
\title {\bf Modeling the Physical Structure of the Low Density Pre-protostellar Core Lynds 1498}
\author {Yancy L. Shirley\altaffilmark{1}}
\affil{NRAO, P.O. Box O, Socorro, NM 87801}
\affil{yshirley@nrao.edu}
\author {Miranda K. Nordhaus\altaffilmark{2}}
\affil{Applied Physics and Astronomy, Rensselaer Polytechnic Institute, Troy, NY 12180}
\affil{nordhm@rpi.edu}
\author{Jana M. Grcevich\altaffilmark{3}}
\affil{University of Wisconsin-Madison, Department of Astronomy, 
       475 North Charter Street, Madison, WI 53706}
\affil{jmgrcevich@students.wisc.edu}
\author {Neal J. Evans II}
\affil{Department of Astronomy, The University of Texas at Austin,
       Austin, Texas 78712--1083}
\affil{nje@astro.as.utexas.edu}
\author{Jonathan M. C. Rawlings}
\affil{Department of Physics and Astronomy, University College London,
        Gower Street, London WC1E 6BT}
\affil{jcr@star.ucl.ac.uk}
\and
\author{Ken'ichi Tatematsu}
\affil{National Astronomical Observatory of Japan, 2-21-1 Osawa, Mitaka
       Tokyo 181-8588, Japan}
\affil{k.tatematsu@nao.ac.jp}
\altaffiltext{1}{Jansky Postdoctoral Fellow at the National Radio Astronomy 
Observatory. The National Radio Astronomy Observatory is a facility of the 
National Science Foundation operated under a cooperative agreement by 
Associated Universities, Inc.}
\altaffiltext{2}{2003 REU Summer Student at NRAO Socorro.}
\altaffiltext{3}{2004 REU Summer Student at NRAO Socorro.}
%%%%%%%%%%%%%%%%%% abstract %%%%%%%%%%%%%%%%%%%%%%%%%%%%%%%%%%%%%%%%
 
\begin{abstract}

Pre-protostellar cores likely represent the incipient stages of low-mass
($\approx 1 \rm{M}_{\odot}$) star formation. 
Lynds 1498 is a pre-protostellar core (PPC) and 
was one of the initial objects toward
which molecular depletion and differentiation was detected.
Despite the considerable scrutiny of
L1498, there has not been a extensive study of the density and temperature
structure as derived from radiative transfer modeling of
dust continuum observations.  We present deep SCUBA observations of L1498
at 850 and 450 \micron , high resolution BEARS maps of the 
\nthp\ $1 \rightarrow 0$ transition, CSO observations of the
\nthp\ $3 \rightarrow 2$ transition, and GBT observations of the
\cts\ $4 \rightarrow 3$ transition.  We also present a comparison of derived properties
between L1498 and nearby PPCs that have been observed at far-infrared and submillimeter wavelengths.
The L1498 continuum emission is modeled using a one-dimensional radiative transfer code that
self-consistently calculates the temperature distribution and calculates
the SED and intensity profiles at 850 and 450 $\mu$m.
We present a more realistic treatment of PPC heating 
which varies the strength of the ISRF, \sisrf , and includes 
attenuation of the ISRF due to dust grains at the outer radius of the core, \av .  
The best-fitted model consists of a Bonner-Ebert sphere with a central density
of $1 - 3 \times 10^4$ cm$^{-3}$, $R_o \approx 0.29$ pc,
$0.5 \leq s_{isrf} \leq 1$, A$_{\rm{v}} \approx 1$ mag, and a nearly isothermal
temperature profile of $\approx 10.5$ K for OH8 opacities.
\cts\ emission shows a central depletion hole while \nthp\ emission is centrally
peaked.  We derive a mean \nthp\ abundance of $4.0 \times 10^{-10}$ relative to H$_2$ that is
consistent with chemical models for a dynamically young yet chemically evolved source. 
The observed depletions of \cts\ and \form , the modest 
\nthp\ abundance, and a 
central density that is an order of magnitude lower than other modeled PPCs 
suggests that L1498 may be a forming PPC.  Our derived temperature and density 
profile will improve modeling of molecular line observations that will
explicate the core's kinematical and chemical state.

\end{abstract}

%%%%%%%%%%%%%%%%%% 1. Main Text %%%%%%%%%%%%%%%%%%%%%%%%%%%%%%%%%%%%%%%%

\section{Introduction}

        Pre-protostellar Cores (PPCs or starless cores) are thought to represent the  
initial stages of low-mass (M $<$ few \msun ) star formation.
PPCs are identified with dense molecular cores ($n \geq 10^4$ \cmv )
which lack evidence for an internal heating source.
Their column density is traced by dust continuum emission (Ward-Thompson et al. 1994;
Ward-Thompson, Motte, \& Andr\'{e} 1999; Shirley et al. 2000; Tafalla et al. 2002) 
and nitrogen-bearing molecular tracers such as
NH$_3$ and N$_2$H$^+$ (Myers \& Benson 1983, Jijina et al. 1999, Caselli et al. 2002,
Crapsi et al. 2005).  It is important to characterize their physical structure
(density, temperature, kinematical, and chemical) to 
constrain the initial conditions of low-mass star formation. 
It is also important to understand the evolution of the 
physical structure of PPCs to trace the evolution of a
nascent protostar.

	Lynds 1498 is a nearby molecular core ($D = 140$ pc) located in the
west-central portion of the Taurus molecular cloud (see Cambr\'{e}sy 1999).    
L1498 was classified as a Pre-protostellar core by its 
submillimeter dust emission and lack of an IRAS detection at 100 \micron\ 
(Ward Thompson et al. 1994).  The relatively weak submillimeter dust emission
indicated that L1498 was a low density core.

        L1498 has been extensively studied with molecular line observations 
and using theoretical chemical modeling 
(Rawlings et al 1992, Taylor et al 1996).  L1498 
was initially identified kinematically as a core potentially 
on the verge of protostellar collapse (Zhou et al. 1994).
Subsequent molecular line observations have shown dramatic evidence for the 
freezing of gas phase molecules onto dust 
grains (e.g., \form , CS, CCS, C$^{18}$O; Wang 1994; Lemme et al. 1995; Kuiper, Langer, \& Velusamy 1996;
Willacy, Langer, \& Velusamy 1998, Tafalla et al. 2004);  
as a result, many molecular species cannot reliably trace the densities in the 
center of the core.  Prominent exceptions are the nitrogen-bearing molecules
\ammonia\ and \nthp\ which resist depletion (Flower, Pineau des For\^{e}ts, \& Walmsley 2005).  
Even though \ammonia\ and \nthp\ remain in the gas phase in
cold, dense molecular cores, the optical depth, gas temperature, and
abundance profile must be reliably determined in order to accurately 
calculate the density structure of the core.

        Optically thin dust emission at submillimeter wavelengths is a good tracer of 
temperature and density since the specific intensity is proportional to the 
integral of density, opacity, and Planck function along a line-of-sight
\begin{equation}
I_{\nu} = \frac{2 \mu m_{\rm{H}} h \nu^3}{c^2} \int_{0}^{\infty} 
\frac{\kappa_{\nu}(s) \, n(s)}{\exp\left(\frac{h\nu}{kT_d(s)}\right) - 1} \; ds
\end{equation}
(Adams 1991, Shirley et al. 2003).  Knowledge of the temperature and density 
structure is essential to a correct interpretation of chemical and kinematic models of L1498.  
Previous dust continuum studies of PPCs used radiative transfer
models to constrain the density and temperature structure of the
cores (e.g. Evans et al. 2001, Zucconi et al. 2001, Stamatellos \& Whitworth 2004).  
The density structures are well characterized by Bonner-Ebert spheres
(hereafter, BE):  pressure bounded, isothermal solutions to the equations of hydrostatic
equilibrium that do not include the effects of magnetic
fields or turbulence.  A general picture of the initial stages of low-mass star formation
involves the evolution of a BE sphere from low central
density to high central density, perhaps retarded by magnetic
pressure, until a central hydrostatic core develops.  If we wish
to understand the beginning of this process, we must identify and
study low density PPCs.

         In this paper we present deep submillimeter maps of
L1498 (\S2.1) with  radiative transfer modeling which includes a more 
sophisticated treatment of the heating of the PPC (\S3). 
The results of dust continuum modeling are discussed in \S4.
A comparison between the derived properties from dust continuum observations
of PPCs that were observed with ISO and SCUBA is discussed in \S4.1.2. 
We also present the highest resolution single-dish \nthp\ and \cts\ 
maps of L1498 (\S2.2, \S2.3) with analysis
of variations in the column density, abundance, and velocity (\S4.2, \S4.3).

\section{Observations}

\subsection{SCUBA Continuum Observations}

\subsubsection{Reduction}

	L1498 was observed simultaneously at 850 and 450 \micron\ 
with SCUBA on the 15 m James Clerk Maxwell Telescope 
during the nights of August 29 and 30, 1998.  A total of fifty jiggle
maps were made towards the source,  each map consisting of four
64-point jiggle maps with 256s of ON-source integration time.  
The total ON-source integration time was 3.55 hours.
Previous spectral line maps showed that L1498 is extended in a SE to
NW direction; therefore, the chop angle was set to a constant position angle of
20\degree\ with a 120\arcsec\ chop throw to chop perpendicular to the major
axis of the core.
Non-azimuthal chopping results in a slight reduction in signal-to-noise in 
the map since a symmetrical chopping pattern does not lie at constant 
airmass.  A preliminary map made in April, 1998 showed that the source was 
detected to the edge of a single 64-point jiggle map field-of-view ($2.3$\arcmin ).  
Therefore, the map was extended using three offset 5-pointing maps, each with
30\arcsec\ spacing (see Figure 1a).  The 
five-pointing maps were centered at (0\arcsec ,0\arcsec ), 
($+$30\arcsec , $-$30\arcsec ), and ($-$30\arcsec ,$+$30\arcsec ) from 
the central coordinates of J$2000.0$ $\alpha =$ $4^h$ $10^m$ $52.5^s$,
$\delta =$ $+25$\degree\ $9$\arcmin\ $55$\arcsec .  
The final maps spans $\pm 140$\arcsec\ in right ascension and declination.

	Each map was reduced using the standard SURF (SCUBA User
Reduction Facility) reduction routines (Jenness \& Lightfoot 1997).
Each 64-point jiggle map was flat-fielded and 
corrected for chop throw, extinction, and 
sky noise.  The 450 \micron\ images were reduced using a set intensity 
scale to ensure consistent identification of improper sky noise subtraction.  
The telescope pointing was checked every hour.  
The largest pointing shift was $2.5$\arcsec; therefore, 
the standard pointing offsets from the five-pointing maps 
were used to shift and add together the individual maps. 

	Skydips at 850 \micron\ and 450 \micron\ were 
performed every hour during both nights.  These skydips are compared to 
the opacity measured every 10 
minutes at 225 GHz ($\tau_{CSO}$) and 350 \micron\ 
from tippers located at the Caltech Submillimeter Observatory.  
The $\tau_{CSO}$ data was only available for the night of August 30.  
Using the relationship derived by Archibald et al. (2002) between $\tau_{CSO}$
and the skydip-determined opacity at 850 \micron\ ($\tau_{850}$),
$\tau_{850} = (3.99 \pm 0.02) (\tau_{CSO} - 0.004 \pm 0.001)$, we found 
excellent agreement between our skydips and the scaled 850 \micron\ opacity 
(Figure 2).  Therefore, we use the the scaled $\tau_{CSO}$ values to 
correct for the 850 \micron\ opacity in individual jiggle maps on the night 
of August 30.  We linearly interpolated between 850 \micron\ skydips for 
the night of August 29.

	Since the sky opacity at 350 \micron\ and 450 \micron\ is very sensitive to 
short term variations in atmospheric precipitable water vapor, we used the tipper opacity 
at 350 \micron\ to monitor the variability of $\tau_{450}$ between the hourly 
skydips.  We found an average ratio of $\tau_{450}$ skydips to $\tau_{350}$ 
of $0.71 \pm\ 0.07$ for both nights.  The 450 \micron\ opacity was corrected 
for each jiggle map using the scaled $\tau_{350}$ opacity interpolated from the tipper
measurements bracketing each jiggle map.

\subsubsection{Images}

	The reduced maps are shown in Figure 1.  L1498 was detected at both
wavelengths with a peak signal-to-noise of $18$ and $13$ for the 850 and 
450 \micron\ maps, respectively.  The contours are spaced by $3\sigma$ for 
both images.  The submillimeter images are characterized by an amazingly flat
intensity plateau.  A 1200 \micron\ continuum map,
smoothed to 20\as\ resolution  (Tafalla et al. 2004), is 
also shown for comparison.   
The FWHM intensity contour of the 1200, 850, and 450 \micron\ maps are
very similar and well fitted by an ellipse with a major axis ($a$) of 197\as ,
minor axis ($b$) of 108\as , and position angle of 122\degree .  The 850 \micron\
and 450 \micron\ ellipses have similar centroids ($+11$\as , $-16$\as ) 
with an uncertainty of $\pm 4$\as\ with respect to the pointing center (\S2.1.1). 
The HWHM radius of the core, defined by the geometric mean of
the major and minor axes ($R_{1/2} = \sqrt{ab}/2$), is 73\as .  While
this radius may be used an a representative size of the core, it is
important to realize that the FWHM contour is asymmetric with an 
aspect ratio ($a/b$) of 1.8.  Furthermore, the 1200 \micron\ and
850 \micron\ maps are non-axisymmetric with a sharper gradient in the
intensity along the northeast edge compared to the southwest edge.

        The ISO map at 170 \micron\ is also shown in Figure 1b
(Ward-Thompson et al. 2002).  This is the shortest wavelength at
which the far-infrared emission has been observed to peak on
the core; ISO 90 \micron\ observations detect a more diffuse
emission peak approximately 3\arcmin\ south of the submillimeter
centroid.  The scale of the 170 \micron\ map is $\pm 300$\as ,
twice the scale in the (sub)-millimeter maps.  The ISO beam
was very large ($\approx 80$\as ) at 170 \micron ; therefore,
the map does not show the non-axisymmetric structure observed on
smaller scales at 850 and 1200 \micron .  The core is elongated,
but with a slightly different position angle on larger scales
(PA $\sim 150$\degree ).  Far-infrared emission is detected on very large scales, 
up to 300\as\ (42,000 AU) from the centroid of the submillimeter emission.

	Azimuthally-averaged, normalized radial profiles of the 850,  
450, and 1200 \micron\ maps were calculated.
The images were re-binned to pixels with half beam spacing
(7\arcsec\ at 850 microns and 3.5\arcsec\ at 450 \micron ) corresponding to
the Nyquist sampling limit in the map.  The centroid was 
chosen to be the average of the 850 and 450 \micron\ centroids 
($+11$\as , $-16$\as ).  The 1200 \micron\ profile was binned at 10\as\ since the
image was smoothed by a 20\as\ Gaussian (Tafalla et al. 2004).
The flat intensity plateau has the same extent in all of the profiles.

	Since L1498 is a very elongated core, azimuthally averaged
profiles will be weighted towards a flatter intensity profile
(see J{\o}rgensen et al. 2002).  This effect may be quantified by considering
the normalized profile within sectors centered on the major and minor axes.
If the sector centered on the major axis is 
chosen with the opening angle given by,
\begin{equation}
\theta_{\rm{maj}} =  2 \tan^{-1}(b/a) \; ,
\end{equation}
where $b/a$ is the ratio of the minor to major axes 
($\theta_{\rm{min}} + \theta_{\rm{maj}} = \pi $), 
then the areas of sectors are equal.  
The azimuthally averaged and sector-averaged profile will be modeled separately in \S3.

\subsubsection{Calibration and the SED}

	Uranus and AFGL618 were observed on August 29 and 30 as flux 
calibrators and for beam profiles at 850 and 450 \micron .  The total flux 
of Uranus was 74.0 Jy at 850 \micron\ and 195.5 Jy at 450 \micron .  The 
apparent diameter of Uranus was 3\farcs 7  effectively making Uranus a 
point source at both wavelengths (see Shirley et al. 2000 for an analysis 
of Uranus' finite size on the beam size).  The flux of AFGL618 was assumed to
be $4.56 \pm 0.17$ Jy/beam at 850 \micron\ and $11.2 \pm 1.4$ Jy/beam at 450 
\micron\ (SCUBA Secondary Calibrator Webpage 
\footnote{www.jach.hawaii.edu/JACpublic/JCMT/Continuumobserving/SCUBA/astronomy/calibration/secondary2004.html}).  
Since we are interested 
in the total flux observed towards L1498, we calibrated the flux in apertures 
of 40\arcsec\ and 120\arcsec .  Flux densities are calculated for an aperture 
diameter, $\theta$, using 
$\Snu(\lambda,\theta) = C(\lambda, \theta) V(\lambda,\theta)
e^{-\tau \sec z}$, 
where $V(\lambda,\theta)$ was the voltage measured at wavelength $\lambda$ 
in an aperture of diameter $\theta$ at airmass $\sec z$ and
$C(\lambda, \theta)$ is the calibration factor. 
The fluxes for L1498 are included in the 
source properties in Table 1.
The uncertainty on the calibration factor was calculated by 
propagating the errors using,
\begin{equation}
\sigma_{\Snu}^{2} = \Snu^{2} \left[ \left( \frac{\sigma_{C}^{2}}{C^{2}} 
\right)_{\rm{Run \: Avg}} 
+ \ \left( \sec^{2}z \: \sigma_{\tau}^{2} \right)_{\rm{Source}}
+ \frac{ \Sigma_{\rm{pixels}} A_i^2 \sigma_{V_i}^2 }{C^2 V^2} 
\right]
\end{equation}
where \Snu\ is the flux density (Jy), $C$
is the calibration factor (Jy/V), $V$ is the total voltage observed 
within the aperture (V), and $A_i$ is the area of a pixel included
within the aperture.

	The SCUBA beam profiles were determined using
observations of Uranus at 850 and 450 \micron .  Uranus was 
observable only at the beginning of our observing shift and only at high
airmass ($\sec z \sim 2$).  The Uranus profile cannot be traced beyond 
60\arcsec\ at 850 \micron\ and beyond 50\arcsec\ at 450 \micron .
Unfortunately, AFGL618 is not strong enough to be used for extended beam 
profiles ($\theta > 25$\as ).  The Uranus profiles are very similar to the 
average Uranus beam profile determined from maps made in April, 1998 
(Shirley et al. 2000).  Since all of our observation were made during 
the morning observing shift (1:30AM HST - 9:30AM HST), the dish should have 
cooled sufficiently such that the beam shape is stable.  No significant 
trends were seen in the April 1998 beam maps observed during the morning 
observing shift; therefore, we feel justified in using the Uranus beam 
profile for all second shift observations.

         The SED was constructed from our SCUBA observations and
a search of the literature (Table 1).  The SED includes previous 
(sub)-millimeter observations by Ward-Thompson et al. (1994)
as well as ISO far-infrared observations at 170 and 200 \micron\
(Ward-Thompson et al. 2002).  The inclusion of the
far-infrared fluxes is extremely important for bracketing the
peak of the SED.

\subsection{\nthp\ Observations }

\subsubsection{BEARS}

Lynds 1498 was mapped with BEARS, a 25-beam double-sideband
SIS receiver array that operates at 3mm (Sunada et al. 2000), 
on April 26, 2004 at the Nobeyama Radio Observatory 45 m telescope.   
The average beamsize within the
array is 17\farcs 8 at 93.17 GHz (Tatematsu et al. 2004); 
however, each beam in the 
array is separated by slightly more than $2 \theta_{\rm{fwhm}}$ (41\farcs 1 ).   
We observed the source using a dithered $4 \times 4$ position-switched
grid on the sky spaced by 20\farcs 55.  The array was rotated by $\pm 90$\degree\ after
each dithered map to change relative beam positions with
respect to the source.  A digital autocorrelator backend 
with $37.8$ kHz resolution ($\approx 0.12$ km/s) was used.
BEARS was tuned to the frequency of the strongest hyperfine
component ($93.1737767$ GHz $J F_1 F = 1 2 3 \rightarrow 0 1 2$,
see Appendix A).  The resulting \nthp\ map has 100 spectra 
separated at nearly full beam resolution and represents the highest
resolution \nthp\ spectrum obtained towards L1498.  

BEARS tuning was consistent across the array with an average
$T_{sys}$ of 220 K.  The intensity was calibrated for the receiver
sideband ratio by observing the bright high-mass star-forming
core, G40.50+2.54, with every beam in the BEARS array and the 
Nobeyama facility single-sideband S100 receiver.  Spectra from the S100 receiver 
were calibrated on the \ta\ scale using the
standard chopper-wheel method (Penzias \& Burrus 1973).
The calibration source was observed twice 
with the array rotated by $90$\degree\ to minimize systematic calibration errors.
The resulting calibration is good to approximately $20$\% .  The spectra
were converted to the \tr\ scale by assuming the BEARS sidelobes are negligible
and using $\eta_{\rm{mb}} = 0.51$ (Tatematsu et al. 2004). 

Pointing observations were performed every hour using the SiO 
$v = 1$, $J = 1 \rightarrow 0$ maser
transition towards NML Tau.  The largest pointing shifts were $6\as $ with average
corrections of $4\as $ and $3\as $ in azimuth and elevation respectively.  The weather 
was clear with winds typically below $6$ m/s during L1498 observations.  
Conditions were good for \nthp\ observations and 
the absolute pointing errors are small compared to the BEARS beamsize.

The data were baselined, summed, calibrated, and interpolated onto a regular spatial grid 
using the \textit{NewStar}\footnote{http://www.nro.ac.jp/~nro45mrt/NEW45m/NewStar/index.html}
software package of the Nobeyama Radio Observatory.
The reduced maps were written out to the group FITS format (Greisen \& Harten 1981)
and read into \textit{AIPS++}.  Fitting of the hyperfine components were performed with
\textit{Glish} scripts in \textit{AIPS++}.  The group FITS data cube was also written to a series 
of ASCII files for calculation of baseline rms and the integrated intensity.

\subsubsection{CSO Observations}

The \nthp\ $J = 3 \rightarrow 2$ transition was observed in position-switched mode
towards the centroid continuum position of Lynds 1498 with the Caltech Submillimeter Observatory
on the night of July 24, 2004.  The 230 GHz double sideband SIS receiver
(Kooi 1992) with a 50 MHz AOS backend were used.  The average $T_{sys}$ was approximately
$430$ K.  The spectral resolution was
147 kHz ($0.15$ km/s) after smoothing to the actual resolution of the
spectrometer (3 channels).  The receiver was tuned to the JPL line catalog
frequency, $279.511701$ GHz, for \nthp\ $J = 3 \rightarrow 2$; this frequency is
within 160 kHz of a blend containing the strongest hyperfine component 
($279.5118631$ GHz, $J F_1 F = 3\, 4\, 5 \rightarrow 2\, 3\, 4$; Luca Dore, priv. comm.).

Pointing and main beam calibration were performed on Venus.
The main beam efficiency was determined to be $\eta_{mb} = 0.60 \pm 0.06$ at
279 GHz for a $26\farcs 8$ beam.  The pointing was consistent with a largest
pointing shift of less than $3$\as\ in both azimuth and zenith angle during the observations.
The spectra were smoothed, baselined, and summed using the \textit{CLASS} software
package (Buisson et al. 2002).  L1498 was observed for a total of $10$ minutes ON-source.  
The final spectrum has a baseline rms of $\sigma(\tr ) = 45$ mK.

\subsection{\cts\ Observations}

The \cts\ $J = 4 \rightarrow 3$ transition at 
23.1229820 GHz\footnote{Frequencies for transitions of \cts\
and its isotopologues are from The Cologne Database for 
Molecular Spectroscopy (M\"{u}ller et al. 2001):
http://www.ph1.uni-koeln.de/vorhersagen/} 
was mapped using the 100m Robert C. Byrd
Green Bank Telescope on the nights of March 14 and 15, 2005.  
The spectrometer was setup to observe
4 IFs with 2 polarizations at 6.125 kHz resolution (0.079 km/s) 
using a single beam of the K-band receiver.  The four IFs
were centered on the \cts\, C$_3$$^{34}$S, $^{13}$CCCS, and C$^{13}$CCS $J = 4 \rightarrow 3$
transitions.
The observations were frequency switched by $+4$ MHz at a frequency of 4 Hz.  
Initially, a pointed map was made using a $7 \times 5$ grid, with 60s integration
time per pointing, spaced at
33\as\ (full beam spacing) centered on the SCUBA dust continuum centroid position.  
Three smaller $3 \times 3$ maps with 16\farcs 5 spacing were centered at the offsets
(0,0), ($+49\farcs 5$,$-16\farcs 5$), and ($-49\farcs 5$,$+16\farcs 5$) with respect to the dust continuum
centroid.

The raw frequency-switched spectra were folded and calibrated on the $T_{A}$ scale
using the \textit{DISH} software package within \textit{AIPS++}.  The data
were then written to ASCII files where opacity corrections were applied ($T_{A}^{'}$)
and RR and LL polarizations were summed with $1/\sigma_{\rm{baseline}}^2$ weighting.
The opacity at $23.1$ GHz was determined using the weather model of R. Maddalena
(private communication) which averages the precipitable H$_2$O measurements from
three nearby towns (Elkins, Hot Springs, and Lewisburg, West Virginia).  The average
zenith opacity was $\tau_{23.1} = 0.049 \pm 0.002$ on March 14 and $\tau_{22.5} = 0.106 \pm 0.002$
on March 15 during the observations.  Final conversion to the $T_{mb}$ scale was achieved 
using position-switched 
scans of the quasars 3C48 and 3C286.  Unfortunately, the aperture efficiency was not properly determined
at $23.1$ GHz on March 14; therefore, we have linearly interpolated from measurements made at $9$ GHz 
(Langston et al. 2004) and $46$ GHz (March 18, 2005) to find $\eta_{a} = 0.54 \pm 0.11$.
The GBT has an active surface which mitigates variations in the aperture
efficiency with elevation and temperature.  We have applied a $20\% $ systematic
uncertainty to account for these effects.  The main beam efficiency is then
$\eta_{mb} = 0.74 \pm 0.15$.  The $T_{mb}$ temperature scale may be compared with
observations on the \tr\ scale if the sidelobe pickup is negligible (see Chapter 8.2.5 of
Rohlfs \& Wilson 2000); this is a reasonable
assumption for the K-band receiver on the GBT.

Pointing and focusing were performed once per hour.  The sky was clear and the
wind was less than $4$ m/s on both March 14 and March 15 during the observations.  
Since the observations were performed
at nighttime, the temperature variation was small ($\Delta T < 1.9$\degree C).  The pointing and
focus were very stable with a largest pointing shift of $7$\as .

\section{Models}

\subsection{1D Continuum Radiative Transfer}

	The dust continuum observations are modeled using the one-dimensional radiative
transfer code \textit{CSDUST3} (Egan, Leung, \& Spagna 1988) that self-consistently
calculates the temperature distribution for a given density distribution,
dust opacity, and ISRF.  \textit{CSDUST3} simulates anisotropic scattering
which may affect the temperature distribution in the outer regions of
the core, a feature that is not present in the radiative transfer
code of Ivezi\'{c}, Nenkova, \& Elitzur (1999).  Once the temperature distribution has been 
calculated,  the intensity profiles at 850 and 450 \micron\ are 
created by convolving the model intensity profiles with the 
measured beamshape and simulating chopping.  The model SED is
also calculated and compared to the observed fluxes.  Simultaneous modeling of
the intensity profiles and SED result in nearly orthogonal constraints
on the shape of the density profile and the total mass for a given opacity choice
(see Shirley et al. 2002).  The modeling procedure for PPCs 
is described in more detail in Evans et al. (2001).

	We shall use a family of BE spheres (Ebert 1955, Bonnor 1956) to attempt to
fit the L1498 observations since BE spheres were used by Evans et al. (2001) to
successfully fit submm continuum observations of other PPCs.  BE sphere
are solutions to the pressure bounded equations of hydrostatic equilibrium
that ignore the effects of magnetic fields and turbulence.
The BE density profile is flat in the inner
regions and asymptotically approaches $r^{-2}$ in the outer regions. 
The size of the flat density plateau decreases as the central
density of the BE sphere increases (density FWHM 
$= R(\frac{1}{2}n_c) \sim 1/\sqrt{n_c}$,
Chandrasekar \& Wares 1949).  The temperature structures are
characterized by decreasing dust temperature from outside to
inside due to extinction of heating radiation from the ISRF.
The temperature gradient from outside to inside increases
for more centrally condensed BE spheres.
Isothermal BE spheres may be parametrized by a single variable, the central
density ($n_c$) for a fixed temperature (10 K).  
Evans et al. (2001) showed that non-isothermal correction 
to the shape of the BE profile was negligible for profiles used to describe PPCs.
Isothermal BE spheres are calculated with central
densities ranging from $\log n_c = 3.5$ to $\log n_c = 7.0$.  
Since L1498 is embedded within a larger molecular cloud with
\av\ $\approx 1 - 3$ mag (Cambr\'{e}sy 1999), 
the power-law portion of the BE profile is not continued below $10^3$ \cmv .

        Since PPCs lack an internal source, the heating is from
the ISRF.  The spectrum of the ISRF is derived from \textit{COBE} data (Black 1994)
with ultraviolet 
wavelengths constructed using an empirical description of the
radiation field that reproduces the ISRF of Draine (1978; see van Dishoeck 1988).
This spectrum  is the same as used in the models of Evans et al. (2001).  
The shape of the ISRF spectrum may be modified by two effects:
the intensity of the ISRF (UV to far-infrared) varies at different
locations throughout the Galaxy; and the short wavelength intensity
of the ISRF (UV to near-infrared) may be modified by dust extinction
(see Figure 3).
The scaling of the intensity is parametrized by \sisrf\ while the
amount of dust extinction at the core's outer radius is parametrized by \av .  We use dust opacities
appropriate for the general ISM (graphite and silicate based grains, 
Draine \& Lee 1984) to modify the ISRF spectrum.
The combination of scaling the ISRF and extinction at the outer 
core radius is a more realistic description of the incident intensity than 
has been used in previous models of PPCs.

        The reprocessing of the ISRF is strongly affected by the
dust opacity within the core.  We model the emission using eight different
opacities models (see Table 2 and Figure 3).  The opacities of 
Ossenkopf \& Henning (1994) are derived from models of
grains that have coagulated for $10^5$ years
and that have acquired varied depths of ice via gas adsorption: 
OH2 no ice mantles (column 2 Table 5 of Ossenkopf \& Henning 1994), OH5 
thin ice mantles (column 5, Table 5),  and OH8 thick ice mantles (column 8, Table 5).  
The opacities of Mathis, Mezger, \& Panagia (1983) are derived from an empirical
fit to dust properties.  We also test the opacities of Weingartner \& Draine (2001) 
which describe silicate and graphite grains, including a population of small grains and
PAHs,  appropriate for the ISM.  The opacities of Pollack (1994) describe
an alternative grain composition based on silicate grains (olivine and orthopyroxene),
iron compounds (troilite and metallic iron), and various organic C compounds.
The mass opacity at 850 \micron , $\kappa_{\nu}(850)$, varies
by more than one order of magnitude between the opacity models (Table 2).
A constant opacity with radius ($R \in [0,R_o]$, $R_o$ is the outer core radius) 
is assumed throughout the core.

\subsection{L1498 Model}

        A total of $770$ radiative transfer models were run that vary
$n_c$, $R_o$, \sisrf , \av , and $\kappa_{\nu}$.
For each model, the reduced $\chi^2_r$, given by
\begin{equation}
\chi^2_r = \sum_{i=1}^{N} \frac{ (x_i - x_i^{obs})^2  
                                   / (\sigma _i^{obs})^2 }
                               {N} \;\;,
\end{equation}
was calculated for the 850 and 450 normalized intensity profiles and
for the SED.  A $\chi^2$ is also calculated for the flux in a
$120$\as\ aperture at 850 \micron .
The best-fitted models are determined by finding the intersection of the
sets of $\chi^2_r$ for each quantity within the typical
range $\chi^2_r \in [0,1]$.  Since the errorbars of the intensity 
profiles include azimuthal variations as well as intensity uncertainties,
and since the errorbars of the SED include an estimate of systematic
uncertainty, it is possible for $\chi^2_r < 1$.

        The first grid of models varies $n_c$ and $R_o$ for 
\sisrf\ $=1$, \av\ $=1$, and OH5 opacities (see top row of Figure 4).  
The best-fitted model
to the azimuthally averaged, normalized intensity profiles is
$n_c = 1 \times 10^4$ \cmv .  
If we re-calculate the $\chi^2_r(I_{850})$
using the sector-averaged, normalized intensity profiles,
then a wider range of central densities fit, 
$n_c = 1^{+2.0}_{-0.5} \times 10^4$ \cmv\ (see dashed profiles in
Figure 5d).  
The best fitted central density does not strongly 
depend on the size of the outer radius;
radii from $30,000$ AU to $60,000$ AU are well fit.
Since emission is clearly detected $100$\as\ from the center
of the SCUBA maps, $R_o$ must be greater than $15,000$ AU.
Furthermore, the ISO 170 \micron\ image (Ward-Thompson et al. 2002)
detects dust emission at radii greater than
$200$\as\ ($28,000$ AU) from the center; therefore, the best-fitted model $R_o$ are
consistent with submillimeter and far-infrared observations.

        Since the SCUBA continuum centroid and 1.2 mm continuum peaks are
different by $31$\as , we have also compared the dust model intensity profiles
to azimuthally averaged intensity profiles centered at the 1.2 mm peak.
The best-fitted model are $n_c = 1 - 3 \times 10^4$ \cmv\ Bonnor-Ebert spheres.
This agreement with the best-fitted models toward the SCUBA centroid position
is a result of L1498's very flat submillimeter intensity profile that does not vary
strongly with different central 
positions within $\approx 30$\as\ of the SCUBA continuum centroid (see Figure 1).

        The second grid of models varies \sisrf , and \av\ for
the $n_c = 1 \times 10^4$ BE sphere, $R_o = 20,000$, $40,000$, and $60,000$ AU,
and the different opacity models.  The \sisrf\ and \av\ are varied on non-linear
grids with values between $[0.2, 10]$ and $[0.1, 10]$ respectively.
The middle and bottom rows of
Figure 4 display $\chi^2_r(SED)$ and $\chi^2_r(S_{850})$ for OH5 dust opacities.
Changing \sisrf\ and \av\ has a negligible effect on the best fitted normalized 
intensity profiles; the $n_c = 1 \times 10^4$ BE sphere is the best fit for
all opacity models.  Previous dust continuum models relied only
upon submillimeter fluxes to constrain the heating properties of the ISRF.
However, $\chi^2_r(S_{850})$ alone do no strongly constrain \sisrf\ and \av\
(bottom row, Figure 5); a wide range of values provide good fits.  When we
include the complete SED, the \sisrf\ is constrained to be
$\leq 2$ for the best-fitted models (middle row, Figure 5).

        The only dust opacity models that provide
satisfactory fits to $I_{850}$, $S_{850}$, and the SED are the Ossenkopf
\& Henning opacities.  None of the model $\chi^2_r(SED)$ for WD3, WD4, Pollack1,
or Draine-Lee opacities are below $6$.  These models underestimate the fluxes
observed at far-infrared through millimeter wavelengths for the $1 \times 10^4$
\cmv\ BE sphere.  This is not surprising since the mass opacity at
850 \micron\ for these models is $3$ to $8$ times smaller than the mass opacity of OH8 dust.  
Furthermore, Ossenkopf \& Henning opacities for grains with ice mantles
(OH5 and OH8) provide better fits to $S_{850}$ and the SED than grains
without ice mantles (OH2).  The OH2 opacities tend to overestimate $S_{850}$ while
underestimating the far-infrared fluxes.  It is encouraging that the dust opacities
for grains with ice mantles provide the best fits since these
models are qualitatively consistent with a physical model of L1498 that accounts for the 
observed gas depletion of species such as CO, H$_2$CO, CS, and CCS.

        The best-fitted models from all of the model grids are
listed in Table 3 and shown in Figure 5.  The models fall into two categories:
OH5 opacities with \sisrf\ $\in [1,2]$ and \av\ $\in [4,8]$; and OH8 and OH5
opacities with \sisrf\ $\in [0.5,1]$ and \av\ $\in [0.5,2]$.  The fluxes and intensity
profiles are well fit at all wavelengths except $1.2$ mm for all of the models
All of the best-fitted models have an outer radius of $60,000$ AU (0.29 pc).  Since a large
outer radius is preferred, it seems unlikely that the high \av\ models are physical.
An optical extinction map of the Taurus molecular cloud limits the large-scale
extinction around L1498 to \av\ $\leq 3$ mag (Cambr\'{e}sy 1999).  Indeed, 
for \av\ $=6$, the column density of material outside $R_o$ would be $N_{H_2} \approx
1 \times 10^{22}$ cm$^{-2}$; this value is larger than the column density derived
from a $1 \times 10^4$ \cmv\ BE sphere with $R_o = 60,000$ AU.  We prefer the low 
\av\ models as the best, most physical fit.

         All of the best-fitted models are characterized by dust temperatures between
$10$ and $11.1$ K (Figure 5a).  The low \av\ OH5 and OH8 models
have temperature gradients of $0.3 < \Delta T_{[0,R_o]} < 0.5$ K.  These are nearly isothermal
temperature profiles compared to previous models of PPCs ($\Delta T_{[0,R_o]} > 1$ K,
e.g. Evans et al. 2001).  The low density of L1498 combined with the inclusion of dust
attenuation at the outer radius results in the nearly isothermal temperature
profiles.

%$T_{iso}$ CALC. $T_{iso}$ $= \frac{h\nu}{k}\left[ 1 + 
%\ln\left(\frac{2h\nu^3M_{env}\kappa_{\nu}}{S_{\nu}c^2D^2}\right) \right]$

\section{Analysis}

\subsection{Derived Properties from Continuum Observations}

\subsubsection{L1498}

	The bolometric luminosity was calculated by integrating under the spectral energy 
distribution of L1498 (Table 1) and is $0.12 \pm 0.02$ \lsun , low for a PPC, 
but not the lowest reported luminosity
(see Table 4).  The bolometric temperature, the temperature of a blackbody with
the same mean frequency as the observed SED (Chen et al. 1995),  
is $13.9 \pm 3.1$ K.  This is one of the lowest bolometric temperatures reported for a PPC
(see Table 4).

	We calculate the mass from the flux at 850 \micron\ in a 120\arcsec\
aperture and assuming a constant density, 
isothermal dust temperature, and optically thin emission using,
\begin{equation}
M_D(T_d) = \frac{\Snu c^2 D^2}{2 h \kappanu \nu^3}
           \left[\exp\left(\frac{h\nu}{kT_d}\right) - 1\right] 
\end{equation}
Based on 1D dust radiative transfer models, a dust temperature 
of $10.5 \pm 0.5$ K is used for L1498.  The resulting 
dust mass, $M_{D}(10.5K) = 0.67 \pm 0.16$ \msun\  for 
an average of OH5 and OH8 opacities ($\overline{\kappa_{\nu}}(850) = 0.020 \pm 0.002$ cm$^{2}$ g$^{-1}$).  
The errors in $M_D$ are propagated from $S_{\nu}$, $T_d$, and $\overline{\kappa_{\nu}}(850)$.  
L1498's dust mass is consistent with and slightly lower than 
previously observed pre-protostellar cores ($\mean{M_D(10.5)} = 0.9 \pm 0.3$ \msun , Table 4).

We also calculate the aperture-averaged column density from the 850 \micron\
fluxes using
\begin{equation}
N_{\rm{H_2}} = \frac{2 S_{\nu} c^2}
                    {h \nu^3 \mu m_{\rm{H}} \kappa_{\nu} \pi \theta_{ap}^2}
               \left[ \exp\left( \frac{h\nu}{kT_d} \right) - 1 \right] \; .
\end{equation}
The column density at the SCUBA continuum centroid is $(1.25 \pm 0.71) \times 10^{22}$ \cmv\
in a $17\farcs 6$ aperture,
where errors are propagated in $S_{\nu}$, $T_d$, and $\overline{\kappa_{\nu}}$.  This aperture
size matches the BEARS beamsize allowing a direct comparison between dust and 
\nthp\ column density (\S4.1.2).  The column density 
corresponds to a visual extinction of \av\ $= 13$ mag (Whittet 2003). 
This extinction is comparable to the peak extinction observed using near-infrared
extinction mapping towards the low-density PPC, Coalsack G2 (\av\ $= 11.5$ mag, Lada et al. 2004).
Modeling of the Coalsack G2 extinction profile indicates a peak density of $\approx 10^4$ \cmv ,
corroborating the low density derived for L1498 from our radiative transfer models.

%Similarly, we can use the best-fitted radiative transfer models to compare column
%densities
%\begin{equation}
%N_{\rm{H_2}}(R) = \frac{16 \ln 2}{\theta_{\rm{fwhm}}^2} 
%                 \int_{-R_{max}}^{R_{max}} \exp \left[ -4 \ln 2 \frac{(R - r)^2}
%                      {\theta_{\rm{fwhm}}^2} \right] 
%		 \int_0^{\sqrt{R_o^2 - r^2}} n(s) \, ds \; r \, dr
%                  + 9.5 \times 10^{20} A_{\rm{v}}\; .
%\end{equation}

	Dust opacities tend to follow a single power-law with frequency at
submillimeter wavelengths ($\kappa_{\nu} \propto \nu^{\beta}$, $\lambda \in 
[350, 1300]$ \micron ).  The observed dust emissivity index, $\beta$, 
for a core observed at 850 and 450 \micron\ is given by,
\begin{equation}
\beta(T_d) = \left[ \frac{\log \left( \frac{S_{450}}{S_{850}} \right) }
                         {\log \left( \frac{850}{450} \right) } \right]
           - \left\{ 3 + \frac{\log\left[ \exp(h\nu_{850}/kT_d) - 1 \over  
                                        \exp(h\nu_{450}/kT_d) - 1 \right]} 
                              {\log\left({850 \over 450}\right)}  \right\} 
           \; ,
\end{equation}
where the term in square brackets is the spectral index between
450 and 850 \micron\ , $\alpha_{450/850}$, and the
term in curly braces is the correction for failure of the 
Rayleigh-Jeans approximation.  Assuming a dust temperature
of $10.5 \pm 0.5$ K, we find $\alpha_{450/850}^{120} = 2.91 \pm 0.61$ and 
$\beta=2.44 \pm 0.62$.
The spectral index is higher than all of the PPCs observed by Shirley et al. (2000)
except for the two cores associated with L1689A.   
The $\beta$ is larger than the theoretical $\beta$ for dust opacities used in modeling 
but is consistent, within errors, with all of the dust opacity models except OH2 and
WD4 (see Table 2).    While OH5 and OH8 opacities
give good fits, within errorbars, to the SED, they consistently underestimate the
flux at 450 \micron\ while matching the 850 \micron\ flux.  Submillimeter
measurements of $\alpha$ and $\beta$ do not discriminate well between opacity
models; radiative transfer modeling of the complete SED is a better discriminator.

\subsubsection{Comparison with Nearby PPCs}

The properties of L1498 may be placed in context by comparing to other PPCs that
have been observed with ISO and SCUBA.  Table 4 lists observed and derived properties of
10 nearby ($D < 200$ pc) sources.  Including the 170 \micron\
ISOPHOT flux gives at least one flux point on the Wien side of the SED, resulting
in a more accurate calculation of \lbol .  Each source SED contains at least
3 flux points (170, 200, and 850 \micron ).
We also analyzed the SCUBA 850 \micron\ images for each source.
All of the SCUBA observations are published except for three cores, L183, L1709A,
and L63 which were obtained from the CADC SCUBA archive.  The reduction procedure
for these cores is identical to that described in \S2.
We have calculated the size, aspect ratio, bolometric luminosity, standard evolutionary
indicators (\tbol\ and \lbolsmm ), and the dust-determined mass from the 850 \micron\
flux.  This is a small sample but probably
representative of nearby PPCs.  The CADC SCUBA archive contains many more
observations of PPCs; however, far-infrared observations are needed to calculate \lbol .

L1498 is the largest nearby PPC in the sample by 50\%\ and nearly twice as large as the average
core size.  L1498 also has one of the lowest masses ($0.67$ \msun ) of the sample. 
This is consistent with the interpretation that L1498 is a low density core with a uniquely flat intensity 
profile.  In contrast, the average nearby PPC is a core with $0.14$ \lsun , $0.8$ \msun ,
and elongated ($\mean{a/b} = 1.6$) with a mean radius of 6000 AU.

PPCs are clearly separated from Class 0 sources on a \tbol\ vs. \lbolsmm\ diagram (Figure 6).  
All of the PPCs have low \tbol\ $< 20$ K and \lbolsmm\ $< 25$ whereas the observed Class 0 sources all 
have \tbol\ $> 20$ K. \tbol\ and \lbolsmm\ for PPCs do not correlate.  This is not
surprising since these evolutionary indicators are strongly influenced by the
strength of the ISRF; this is an environmental effect not directly related to the
evolutionary state of the core.  A better scheme for characterizing the evolutionary
state of PPCs would be the comparison of the dynamical maturity determined from
radiative transfer models ($n_c$) and
the chemical state of molecules that are abundant at different times.  For
instance, comparisons of [CCS]/[\nthp ], comparisons of [N$_2$D$^+$]/[\nthp ],
and the depletion fraction of, $f_D$, of 
C$^{18}$O and \form\ are promising candidates.  The evolutionary state of
L1498 is discussed in \S4.3.

\subsection{Derived Properties from \nthp\ Observations}

The seven hyperfine lines of \nthp\ $J = 1 \rightarrow 0$ were well detected over 
an extent of $\pm 80$\as\ in the map (Figure 1f). 
The peak \ta\ in the map is coincident with the continuum centroid (Figure 7a). 
We can estimate the effective density needed to excite an easily observable 1 K
\nthp\ line (see Table 1 of Evans 1999).  The critical density of the $J = 1 \rightarrow 0$
is $n_{crit} = 3.8 \times 10^5$ \cmv\ (collision rates determined from Turner 1995).  
By analogy with \hcop , the resulting effective density
(corrected for optically thin hyperfine structure) is $n_{eff} \approx 2 \times 10^4$ \cmv , 
consistent with the derived central density.

The integrated intensity is calculated by 
integrating over the seven hyperfine components.  The $I(\tr )$ at the continuum centroid is also 
the peak in the \nthp\ map, $I(\tr ) = 3.92 \pm 0.48$ K km/s.  The FWHM
integrated intensity contour agrees very well with the FWHM continuum contour
in size and position angle.  The \nthp\ emission appears to trace the general
shape of the continuum emission, but does not peak as sharply along the north-eastern
ridge as is seen in the 850 \micron\ and, especially, in the 1200 \micron\ images.

The hyperfine structure was fitted with seven Gaussian functions, constrained by 
the hyperfine transition velocities (Appendix A), using the profile fitter
task in \textit{AIPS++}.
The average linewidth was calculated to be $0.22 \pm 0.02$ km/s over the map
using the isolated hyperfine component ($JF_1F = 1\, 0\, 1 \rightarrow 0\, 1\, 2$).  
This linewidth is slightly smaller
than the linewidth of $0.28$ km/s found by Caselli et al. (2002) using the FCRAO. 
The non-thermal contribution to the linewidth is $\Delta v_{nt} = 0.13$ km/s.
A plot of the linewidth versus radius (Figure 7b) indicates that 
the linewidth does not vary significantly out to $15,000$ AU ($\approx 110$\as ).  
Therefore, the average gas motions in L1498 have low turbulence and are similar throughout
the core.  This result agrees with the $\Delta v$-$r$ relationship towards other
PPCs in Taurus observed with BEARS (Tatematsu et al. 2004).

	We calculate the virial mass using the Doppler linewidth and assuming an isothermal dust 
temperature,
\begin{equation}
M_{vir} = { {(6 R_{FLAT} + 9 R_{1/2})} \over 8 G \ln{2} } \left[ {{k T_{iso} 
\over \mu m_{H}} + {\Delta v^2 \over 8 \ln{2}} -{ k T_{iso} 
\over m_{amu} m_{H}}} \right] \; .
\end{equation}  
For a more accurate calculation, we split the radial profile into 
two sections.  The first corresponds to $R \leq\ R_{FLAT}$ (the radius where $n = 0.9 n_c$) 
which is approximated by a constant density.  
For a $1 \times 10^4$ \cmv\ BE sphere, $R_{FLAT}$ is  $520$ AU.
Within the second section, $R_{FLAT} \leq\ R \leq\ R_{1/2}$, the 
density can be approximated by an inverse power law with $r^{-2}$
($R_{1/2} = 10200$ AU, \S2.1.2).  
The term in square brackets converts the observed \nthp\ linewidth to a 3D velocity dispersion 
(Shirley et al. 2002).  Using the $\Delta v$ determined above, 
we find $M_{vir}=0.84 \pm 0.21$ \msun\ with errors propagated in 
$R_{1/2}$, $T_{iso}$, and $\Delta v$.  This mass agrees
with the mass determined from dust continuum emission (\S4.1.1)
indicating that L1498 is likely a bound core and that the best-fitted Ossenkopf \& Henning
dust opacities are appropriate.

The \nthp\ $J = 1 \rightarrow 0$ transition is observed to have moderate optical
depths towards low-mass cores (e.g. Tatematsu et al. 2004, Crapsi et al. 2005); 
therefore, we must calculate
the optical depth at each position in the map to derive the column
density.  For a position-switched observations, theoretically we observe the quantity 
$T_{R} = ( J_{\nu}(T_{ex}) - J_{\nu}(T_{bg}) )[1 - \exp(-\tau)]$,
where $J_{\nu}(T) = (h\nu/k) / [\exp(h\nu/kT) - 1]$.  This $T_{R}$ is
related to the actual observed quantity, \ta ,  through source-beam coupling and
through the efficiency of the antenna (including spillover and atmospheric
attenuation, see Kutner \& Ulich 1981).  Since the hyperfine lines lie close
together in frequency ($\Delta \nu < 4.65$ MHz), we assume that the proportionalities
for each line are constant (Keto et al. 2004).
Therefore, we find the optical depth in the strongest hyperfine component 
($\tau_m$, $JF_1F = 1\, 2\, 3 \rightarrow 
1\, 1\, 2$) by taking ratios of the observed antenna temperatures of the $i^{\rm{th}}$
hyperfine component to the strongest hyperfine component 
\begin{equation}
\frac{(T_A^*)_i}{(T_A^*)_m} = \frac{1 - e^{-R_i \tau_m}}{1 - e^{-\tau_m}} \; ,
\end{equation}
where $R_i$ is the ratio of relative strengths of the hyperfine lines to
the strongest line.  $R_i$ is calculated in Appendix A.

The average optical depth in the strongest hyperfine line, $\mean{\tau_m}$, 
was calculated using four of the six observed line ratios and Equation 9.  
Two of the lines ($JF_1F = 1\, 1\, 2 \rightarrow 0\, 1\, 2$ and 
$1\, 1\, 0 \rightarrow 0\, 1\, 1$)
consistently gave lower $\mean{\tau_m}$ and higher $\mean{\tau_m}$
by factors of 0.4 and 2.0 respectively.  This may be due to anomalous hyperfine
excitation, an effect that has been observed towards other PPCs (Caselli et al. 1995,
Turner 2001).  An optical depth was determined for 47 positions with suitable signal-to-noise.
The optical depth is not affected by line blending since the linewidth is $0.22$ km/s
(see Appendix A).  Since, each hyperfine 
line has a different optical depth, then correcting the column density using
only $\tau_m$ would overestimate the column density.  We define the
average tau by averaging the 
theoretical hyperfine ratios for the seven observed transitions 
$\mean{\tau} = \frac{1}{7} \sum  
R_i \tau_m = \frac{27}{49} \tau_m $.
The average optical depth in the map is $\mean{\tau}_{map} = 0.83 \pm 0.54$, indicating that the
\nthp\ $J = 1 \rightarrow 0$ transition is moderately optically thick over most of the map.
As an example, the SCUBA continuum centroid position has an optical depth of $\mean{\tau} = 1.11 \pm 0.61$,
resulting in a correction for optical depth to the column density of 
$\mean{\tau}/(1 - e^{-\mean{\tau}}) = 1.66^{+0.43}_{-0.38}$.

The total column density for \nthp\ is
(see Goldsmith \& Langer 1999)\footnote{This derivation of the correction for an optically
thick column density should not be confused with estimates based on the total optical depth, 
$\tau_{tot} = \frac{27}{7}\tau_m$ (e.g. Caselli et al. 2002b).}
\begin{equation}
N_{\rm{N_2H^+}} = \frac{3 k Q(T_{ex}) \exp(E_u/kT_{ex})}{8 \pi^3 \nu \mu^2 J_u}
\frac{\mean{\tau}}{1 - e^{-\mean{\tau}}}  \int \frac{T_A^*}{\eta_{\rm{mb}}} dv \; ,
\end{equation}
where $Q(T_{ex})$ is the rotational partition function and $\mu = 3.4$ D (Green et al. 1974)
is the dipole moment.  We assume an excitation temperature of $10 \pm 5$ K. 
Errors in $N_{\rm{N_2H^+}}$ are propagated for $T_{ex}$, $\mean{\tau}$, 
and $I(\tr )$.  The peak column density is $(8.3 \pm 3.4) \times 10^{12}$ cm$^{-2}$
with an average column density of $(3.5 \pm 2.1) \times 10^{12}$ cm$^{-2}$.  The peak
column density agrees remarkably well with the calculation of Caselli et al. (2002a) for
lower resolution FCRAO 15 m \nthp\ observations towards L1498 ($(8 \pm 4) 
\times 10^{12}$ cm$^{-2}$)
and the intermediate resolution IRAM 30 m observations of Crapsi et al al. (2005;
$(7.1 \pm 0.7) \times 10^{12}$ cm$^{-2}$).

We calculate the abundance, $X_{\rm{N_2H^+}}(R) =  N_{\rm{N_2H^+}}(R) / N_{\rm{H_2}}(R)$,
by comparing the \nthp\ column density to the beam averaged column
density determined from the 850 \micron\ dust continuum map (Equation 6).
The average column density in the map is 
$\mean{X_{\rm{N_2H^+}}}_{map} = (4.0 \pm 3.7) \times 10^{-10}$ relative to H$_2$.
This column density agrees very well with the early-time abundances of the 
coupled dynamical-chemical models of Lee et al. (2004) for quasi-statically evolving 
PPCs (\S4.3). 
The abundance show no significant variations with radius out to $15,000$ AU,
although the errors on the abundance are sizable (Figure 7d).  
Radiative transfer models
of the \nthp\ emission should assume a constant abundance of $\approx 4 \times 10^{-10}$.
%The abundance as a function of radius to shown in Figure ZZ.  Errors in 
%$\kappa_{\nu}$ have been ignored for the errorbars because we do not consider opacity changes
%with radius; since $N_{\rm{H_2}} \propto 1/\kappa_{\nu}$, the abundance scales linearly with dust opacity.

The \nthp\ $J=3 \rightarrow 2$ transition was not detected towards the continuum centroid
position with a $1\sigma(\tr )$ baseline rms of $45$ mK (Figure 7c);
however,   
Crapsi et al. (2005) report a detection with the IRAM 30 m 
toward the 1.2 mm continuum peak position (\ta\ $= 380 \pm 53$ mK;  
see inset in Figure 7c). 
%It is not surprising that the $J=3 \rightarrow 2$ transition would be collisionally excited in 
%these cores and not towards the low density core L1498; the effective 1 K excitation density is 
%$n_{eff} \approx 1 \times 10^5$ \cmv .  
This is puzzling since the
$J=1 \rightarrow 0$ BEARS map shows no enhancement toward that position. 
It is possible that the 1.2 mm continuum peak represents the true density peak of
L1498.  A high-resolution, oversampled 
map of the \nthp\ $J=3 \rightarrow 2$ emission is needed to resolve this discrepancy.
The exact location of the density peak will be important for molecular line radiative
transfer modeling, but does not affect the conclusions from dust continuum modeling (\S 3.2).

The L1498 \nthp\ observations will be modeled, along with other PPCs, in a future paper.

\subsection{Derived Properties from \cts\ Observations}

The \cts\ $J = 4 \rightarrow 3$ line was strongly detected toward L1498 (Figure 8) and 
the emission follows the same double-peaked pattern
as previous observations of CS and CCS (Figure 1 of Willacy, Langer, \& Velusamy 1998).  
The line intensity is strongest in the southeastern peak (\tr\ $= 1.64 \pm 0.35$ K) at 
J$2000.0$ $\alpha =$ $4^h$ $10^m$ $56.9^s$, $\delta =$ $+25$\degree\ $9$\arcmin\ $22\farcs 5$. 
Since the \cts\ emission has an intensity saddle at the SCUBA dust continuum centroid, 
it is another excellent example of strong chemical differentiation in L1498.
The southeastern \cts\ peak does not exactly correspond to the Willacy, Langer, \& Velusamy CS or CCS
peak, but instead lies between them and closer to the CS peak.  Combination of the GBT
observations with higher resolution observations (e.g., VLA) would improve the peak position
and may confirm slight chemical differentiation
between all of the observed sulfur-carbon chain molecules.

The \cts\ $J = 4 \rightarrow 3$ linewidth was measured using the Gaussian fitting routine is
\textit{AIPS++} toward 42 positions in the map with detections greater than $3 \sigma$.  
The average linewidth is $\Delta v = 0.21 \pm 0.03$ km/s,
consistent with the \nthp\ linewidth.  No significant variation of $\Delta v$ is also seen for \cts\
with radius (Figure 8b).

The column density of \cts\ was estimated using the same method as for \nthp\ (Equation 10). 
%except that a correction factor for $T_{cmb}$ in the frequency-switched observations was applied.
We estimate the optical depth of the \cts\ $J = 4 \rightarrow 3$  with 10 minute observations
at the southeastern peak position of three isotopologues.  
The detection of C$_3$$^{34}$S $J = 4 \rightarrow 3$
(Figure 8c; $T_A^{\prime}($C$_3$S$)/T_A^{\prime}($C$_3$$^{34}$S$) = 19.7 \pm 3.1$) 
and lack of a significant detection ($> 3\sigma$) of $^{13}$CCCS or 
C$^{13}$CCS $J = 4 \rightarrow 3$, indicates, for the standard ISM isotope
ratios of [$^{32}$S]/[$^{34}$S] $= 22$ and [$^{12}$C]/[$^{13}$C] $= 77$ 
(Wilson \& Rood 1994), that the \cts\ emission is consistent
with being optically thin.  
The peak column density is $ N_{\rm{C}_3\rm{S}} = (5.0 \pm 1.8) \times 10^{12}$ cm$^{-2}$ and 
is less than a factor of 2 less than the peak CCS column density determined by 
Wolkovitch et al. (1997) ($8.7 \times 10^{12}$ cm$^{-2}$).

The abundance of \cts\ was calculated by comparing the \cts\ column density to the
column density derived from the 850 \micron\ map in 33\as\ apertures.
Since the \cts\ emission is very asymmetrical, we plot the 
abundance versus radius in the major-axis sector (\S2.1.2) in Figure 8d.
Negative radii refer to positions SE of the dust continuum centroid.  The depletion of
\cts\ is clearly visible with enhanced abundances by factors of $4.8$ and $2.3$
at the two \cts\ intensity peaks.
The abundance is largest in the SE peak ($X_{\rm{C}_3\rm{S}} = (4.5 \pm 2.4) \times 10^{-10}$).
The abundance peaks are symmetrically located at a radius of 9500 AU from the dust
continuum centroid position.  This radius set a limit for the depletion radius
of \cts , corresponding to a density of $0.8 - 1.5 \times 10^4$ \cmv\ for the best-fitted
BE spheres.  Since the abundance pattern of \cts\ is similar to smaller 
sulfur-carbon chain molecules (e.g., CS and C$_2$S), the southeastern \cts\ peak of
L1498 would be an excellent site to search for larger sulfur-carbon chain molecules
such as C$_4$S and C$_5$S (Gordon et al. 2001).

\subsection{Interpretation of Density Structure}

Radiative transfer modeling of submillimeter images combined with far-infrared and millimeter
photometry indicate that L1498 is a low density, nearly isothermal PPC, with a low strength of
the ISRF and dust opacities appropriate for coagulated, icy grains.  
The central density of $1 - 3
\times 10^4$ \cmv\ is the among lowest central densities reported for PPCs, similar to 
Coalsack G2 (Lada et al. 2004).  This density range agrees with the densities determined
by Langer \& Willacy (2001) from analysis of ISO observations of L1498 ($1.2 - 5.5 \times 10^4$ \cmv\
with lower densities more appropriate for larger core radii).  
However, our modeling results differ substantially from
the analytical model of Tafalla et al. (2004) which fits the radial profile of L1498
at 1.2 mm with a function of the form $n(r) = n_c / [1 + (r/R_{1/2})^{p}]$
and assumes an isothermal dust temperature profile of $T_d(r) = 10$ K.  Tafalla et al.
find a central density of $9.5 \times 10^4$ \cmv , three times larger than the maximal BE
central density in our models.  There are a few discrepancies that may account
for this difference.  Tafalla assume a dust opacity of $5 \times 10^{-3}$ cm$^{2}$g$^{-1}$
at 1.2 mm that is a factor of $2$ smaller than OH5 opacities ($1.02 \times 10^{-3}$ cm$^{2}$g$^{-1}$).
Since the column density is inversely proportional to the opacity, the Tafalla et al. opacity
would result in a factor of 2 larger central density.  This does not account for
the full difference between the results.  The Tafalla et al. centroid position is located at the
peak of the non-axisymmetry which is much more pronounced at 1.2 mm than at submillimeter
wavelengths.  Tafalla et al. calculate a column density of
$3$ - $4 \times 10^{22}$ cm$^{-2}$, a factor of 3 higher than our column density determined
at 850 \micron .  For all of the best-fitted radiative transfer models, the 1.2 mm flux is 
always underestimated while the other six fluxes (170, 200, 450, 850, 1100, and 1300 \micron )
are fit within the errorbars.  The
1.2 mm flux point appears to be anomalously high.  It is possible that calibration uncertainties may 
account for the remaining difference.

The low density implies that L1498 is dynamically young.  Indeed, L1498 may be in
a stable hydrostatic state.  One measure of stability of BE spheres is the density
gradient between inside and outside of the core.  If the density contrast exceeds $14.3$, 
then the BE sphere is in an unstable equilibrium (see Foster \& Chevalier 1993).  
A lower density limit of $10^3$ \cmv\ was assumed for the core in the radiative transfer models.  
This lower limit was chosen to be larger than the average density in a molecular cloud ($10^2$ \cmv\ ) 
since L1498 is situated within an extended condensation of $1$ to $3$ \av\ 
in the Taurus-Auriga molecular cloud (see Cambr\'{e}sy 1999).
The average solution ($1 \times 10^4$ \cmv ) is stable, but this
results depends sensitively on the density at the outer radius.
For the maximal central density ($3 \times 10^4$ \cmv ), the density contrast is $30$ and
the core is unstable.  Unfortunately, we cannot strongly constrain the density contrast since
the submillimeter observations are insensitive to structure outside the chop distance 
(120\as\ or 16,800 AU).

A more quantitative way to test the stability is to
directly compare the Jeans mass with the mass of the BE sphere.
The Jeans Mass for $T = 10.5$ K and $n = 10^4$ \cmv\ is 
$M_{\rm{J}} \approx 18 T^{3/2} n^{-1/2} = 5.7$ \msun\ (Spitzer 1978), while 
the $1 \times 10^4$ \cmv\ BE sphere has a mass of 
$M_{BE} \approx 1.15 (T/10 \rm{K}) (P_o / 10^5 \rm{K} \rm{cm}^{-3})^{-1/2} = 3.6$ \msun\ 
(Bonnor 1956).  The BE mass is within a factor of 2 of the Jeans mass, but $M_{BE} < M_{J}$.
If we use the $3 \times 10^4$ \cmv\ BE sphere, then the mass is comparable to
the Jeans mass $M_{\rm{J}} = 3.3$ \msun .
This mass comparison is consistent with L1498 being a core that is marginally stable to
gravitational collapse.

While L1498 may be dynamically unevolved, it appears to be chemically evolved.
Species such as \form , CS, CCS, and \cts\ (e.g., Willacy et al. 1998, Tafalla et al. 2004, 
Young et al. 2004, \S4.3) are depleted towards the center of the core.  We can estimate the 
depletion timescale 
from the rate equation given in Rawlings et al. (1992).  For a depletion fraction, $f_{\rm{D}}$, 
the adsorption timescale of a neutral molecular species, $i$, is given by,
\begin{equation}
t_{\rm{adsorp}}(i) = \frac{1.38 \times 10^9 \; \ln(1/f_{\rm{D}}) m_i^{1/2}}{n \, S_i \, T^{1/2}} \; \rm{yrs.},
\end{equation}
where $m_i$ is the molecular mass in a.m.u. and $S_i$ is the sticking coefficient
(also see Charnley, Rodgers, \& Ehrenfreund 2001).  
For example, \form\ has been observed to 
be severely depleted towards L1498 over a large spatial
extent (Wang 1994; Young et al. 2004).  The adsorption timescale is 
$t_{\rm{adsorp}}(\rm{H}_2\rm{CO}) \approx 5.5 \times 10^5$ years for $f_{\rm{D}} = 1/10$, 
$S_i = 1$, and $T = 10.5$ K in the absence of other formation or destruction
mechanism.  In reality, many chemical effects may affect this timescale.
Two of the first molecules to be depleted are CCS and \cts .  
Comparisons with detailed coupled dynamical-chemical 
models indicate that the CCS becomes depleted on timescales of a few $10^5$ years (e.g., Li et al. 2002,
Aikawa et al. 2003, Lee et al. 2004).
%Since $t_{\rm{adsorp}}$ scales as $m_i^{1/2}$, then a lower limit to the chemical
%age of the core is set by the heaviest species with observed depletion.  The adsorption timescale
%is 1.8 times longer for CCS than for \form .  [JONATHAN - the CCS timescale seems long ?]

A chemical age of a few $10^5$ years is consistent with the observed abundance of
\nthp\ (Aikawa et al. 2003, Lee et al. 2004).  \nthp\ is considered a ``late time'' species
since it is primarily destroyed in the gas phase by CO and its abundance increases
after CO is significantly adsorbed (although dissociative recombination
may also be important at low densities, see Geppert et al. 2004).  Detailed chemical
models indicate that this abundance increase occurs after $\approx 10^5$
years (e.g. Aikawa et al. 2003).  Clearly, L1498 must be have been in a stable or very 
slowly evolving configuration for at least $10^5$ years to display the extreme molecular 
depletion and modest \nthp\ abundance that is observed.

The dust opacities that best-fitted the L1498 SED were Ossenkopf \& Henning opacities
(1994) for grains that had coagulated for $10^5$ years and accreted ice mantles (OH5 \& OH8).
These opacities have also successfully fit the SEDs of low-mass embedded protostars
(e.g., Shirley et al. 2002) and high-mass embedded protostars (e.g., Mueller et al. 2002). 
The OH opacities are qualitatively consistent with the observed molecular depletion and
the chemical timescale for L1498; however, the coagulation simulation of OH5 and OH8 assumes a density
of n$_{\rm{H}_2} = 10^6$ \cmv\ sustained for $10^5$ years (see Ossenkopf 1993).  
This density is substantially higher than the density derived from dust continuum models; therefore,
the OH5 and OH8 opacities may not be based on an accurate representation of fluffy aggregation
of grains in low density PPCs.

How do you create a dynamically young yet chemically evolved core?  If L1498 is
in an unstable equilibrium, it cannot have been in that state for a long period of
time since the free-fall time for a $1 - 3 \times 10^4$ \cmv\ BE sphere, 
$t_{ff} = \sqrt{3\pi / 8 G \mu m_{\rm{H}} n}$ (Spitzer 1978), is approximately
$3 - 1.7 \times 10^5$ years.  Clearly, L1498 has not yet collapsed and displays
chemical differentiation that requires a similar timescale. 
L1498 must have been static or slowly collapsing for more than $10^5$ years.
While magnetic fields may also play a significant role in 
cloud support and evolution for dynamically nascent PPCs (e.g. Mouschovias \& Spitzer 1976, 
Li et al. 2002), in the case of L1498, they are not necessary for support of the cloud
(since $M_{BE} < \rm{or} \approx M_J$). 
Nevertheless, we can estimate the strength of the B-field required for 
equipartition of magnetic, gravitational,
and kinetic energy.  The strength of this B-field is given by
$B \approx 0.51 n^{1/2} \Delta v_{nt}(\rm{km/s}) = 8.7$ $\mu$G (Lada et al. 2004).
This is a small magnetic field compared with Zeeman measurements of the CCS
$J_{N} = 3_2 \rightarrow 2_1$ transition which provide a rough estimate of
the line-of-sight field strength of $48 \pm 31$ $\mu$G (Levin et al. 2001).  
If L1498 becomes massive enough to be unstable to gravitational collapse, then
the magnetic field appears to be strong enough to support L1498, in which
case, the timescale for collapse is controlled by the ambi-polar diffusion
timescale until the core becomes magnetically super-critical.

Recent molecular line studies of PPCs in the Taurus-Auriga molecular cloud have
identified cores (e.g. L1521B and L1521E)
that appear to be dynamically young and chemically young
(Hirota et al. 2002; Tafalla \& Santiago 2004;
Hirota, Maezawa, \& Yamamoto 2004).  The signature of chemical youth is strongly peaked 
sulfur-bearing carbon chain molecules (e.g., CCS and C$_3$S), 
weak \ammonia\ and \nthp\ emission, and low molecular depletion of C$^{18}$O.  
Confirmation of the chemical and dynamical youth of these
cores requires a detailed comparison between submillimeter dust continuum emission
and high-resolution observations of CCS or C$_3$S.
These objects, along with L1498, L1512 ($n_c = 10^5$ \cmv ), and L1544 ($n_c = 10^6$ \cmv )
may represent an evolutionary
sequence in Taurus-Auriga where chemical maturity is reached first and then dynamical evolution
occurs (see Figure 9).

An important caveat is that our 1D modeling has ignored the observed non-axisymmetry in L1498.  
The dust continuum emission peaks along the north-east ridge and not at the FWHM centroid.
This effect is most pronounced in the 1.2mm but is also detected at 850 \micron .
Other PPCs (e.g., L1544 and L63) also display clear non-axisymmetries.
The spherical BE model is limited and unable to describe this structure. 
A full 3D radiative transfer modeling is needed to probe the structure of
the non-axisymmetry (e.g. Doty et al. 2004).  Also, since the 90 \micron\
ISO image (Ward-Thompson et al. 2002)
shows a clear gradient across the image, there is evidence that the
ISRF asymmetrically heats L1498 (also see Langer \& Willacy 2001).  
3D models could explore triaxial, BE-like
model with a spatially varying ISRF compared to fully non-axisymmetric density
distributions (see Gon\c{c}alves, Galli, \& Walmsley 2004).  
Encouragingly, a comparison between 3D models and 1D models of
L1544 indicate that the 1D models are an acceptable fit to the average physical
structure (Doty et al. 2004)

%	Time scales: how long to deplete CCS, H2CO etc.  Compare to dynamical
%models of BE spheres - Foster and Chev. 1993.  What is free-fall time
%for $10^4$?  Magnetic support account for factor of 2 in density?
%Seems like a young PPC based on central density structure but is
%chemically evolved!  Is this a stable BE sphere (density contrast
%less than 14.3 from inside to outside - this depends on what you
%choose for the outer radius density: $10^3$ seems reasonable but may
%be lower).  Non-axisymmetry - asymmetrical heating from ISRF?  Some magnetic
%models are non-axisymmetric - see Galli and Li papers on isothermal 
%toroids.  Need 3D models.  However, we can find an empirical density
%structure that sufficiently described the structure in 1D with
%uncertainty in the central density of $\approx 2 \pm 1 \times 10^4$ \cmv .  

\section{Conclusion}

We have presented deep SCUBA observations, BEARS and CSO \nthp\ observations,
and GBT \cts\ observations 
of the Pre-protostellar core, Lynds 1498.  Radiative transfer modeling of
the submillimeter intensity profiles and the complete SED were
performed.  Our main conclusions are as follows.

L1498 is characterized by a low density Bonnor-Ebert sphere with central density of
$1 - 3 \times 10^4$ \cmv\ and outer radius greater than $40,000$ AU.
The SED is best-fitted with Ossenkopf \& Henning opacities for coagulated grains with
thick and thin ice mantles (OH8 and OH5 opacities).  A more realistic treatment
of the ISRF was used that includes variations in the strength of the incident specific intensity
and extinction due to dust grains at the outer radius of the core.  
L1498 is a potentially stable ($M_{BE} < M_{J}$), magnetically
sub-critical core that appears to be dynamically young.  These results do not 
change if the 1.2 mm continuum peak is used for the radial intensity profiles instead of
the SCUBA continuum centroid.

Observations of the depletion of species such as \cts\ and \form\ indicate L1498 is 
chemically evolved (t $> 10^5$ years).  
The modest \nthp\ abundance is also consistent with a chemically evolved source.
The best-fitted Ossenkopf \& Henning opacities are
consistent with this timescale, although the Ossenkopf coagulation models
assume a higher density than is observed toward L1498.
The \nthp\ and \cts\ linewidth do not vary significantly 
with radius out to 15,000 AU.  The \nthp\ abundance also does not vary significantly
with radius, while the \cts\ abundance indicates significant depletion toward the 
center of L1498.

Comparisons of L1498 with nearby PPCs that have also been observed at far-infrared
and submillimeter wavelengths indicate that L1498 is larger than average, less luminous than
average, and less massive than average.  The standard evolutionary indicators used for low-mass
protostars, \tbol\ and \lbolsmm , do not correlate for PPCs.  All of the PPCs in this
sample are characterized by \tbol\ $< 20$ K and \lbolsmm\ $< 25$.

The temperature and density structure derived from 1D models may now be used in more
realistic models of molecular line emission.  Future studies of PPCs require high resolution 
far-infrared observations coupled with 3D radiative transfer modeling of multiple wavelengths.  
Instruments such as HAWC on SOFIA and the launch of HERSCHEL will provide the needed high resolution 
far-infrared observations.  

\appendix

\section{Hyperfine Levels of \nthp\ $J = 1 \rightarrow 0$}

Since every nucleus in the \nthp\ molecules has non-zero
nuclear spin, \nthp\ has a very rich hyperfine spectrum that
can be exploited to probe the physical conditions in 
star-forming regions.  In this appendix, we summarize the
hyperfine splitting of the ground state rotational transition of
\nthp .  

The hyperfine splitting (in order of importance) is due to:
(1) an electric quadrupole interaction due to the outer nitrogen; (2) an electric quadrupole
interaction due to the inner nitrogen; (3) a magnetic dipole interaction due to the outer
nitrogen; (4) a magnetic dipole
interaction due to the inner nitrogen; 
(5) a magnetic dipole interaction due to the hydrogen nucleus; and (6) spin-spin interactions
between the nuclei.  The final two effects are ignored
in calculations of hyperfine splitting for astronomical sources since
the observed linewidth is $> 0.1$ km/s ($> 31$ kHz) and the hyperfine coupling constants
are small (see Caselli, Myers, \& Thaddeus 1995).

There are 15 electric dipole allowed transitions for \nthp\ $J = 1 \rightarrow 0$
in the $|J F_1 F\rangle$ basis
($\Delta J = -1$, $\Delta F_1 = 0,\pm 1$, $\Delta F = 0,\pm 1$, $0 \not\rightarrow 0$).
The 15 transitions are listed in Table 5; however, since in practice the $J = 0$ level is not split, 
degenerate transitions are collected together into the 7 observed transitions.  
We adopt the standard notation from Caselli, Myers, \& Thaddeus for the transition labels 
(1995).

It is convenient to calculate the ratio of relative strengths, $R_i$, of the hyperfine levels
for calculations of optical depth and column density (\S4.2).
The relative strengths, $s_i$, are determined using irreducible
tensor methods (Gordy \& Cook 1984, Chapter 15)
and are defined such that the sum of the relative 
strength of the electric dipole allowed 
transitions are equal to one (see Rudolph 1968).
The ratio of relative strengths for the \nthp\ $J = 1 \rightarrow 0$ transitions in the
$|J F_1 F\rangle$ basis are given by
\begin{equation}
R_i(1 F_1^\prime F^\prime \rightarrow\ 0 F_1 F) =  
\sum_{\stackrel{\rm{allowed}}{\rm{F^\prime \rightarrow F}}}  \frac{3}{7}
\left\{ \begin{array}{ccc} 
1   & F_1^\prime & 1 \\ 
1   & 0          & F_1 
\end{array}
\right\} ^2 
\left\{ \begin{array}{ccc} 
1   & F^\prime & F_1^\prime \\ 
1   & F_1      & F
\end{array}
\right\} ^2 
\prod_{F_i = F_1^\prime, F_1, F^\prime, F} (2F_i + 1) \, ,
\end{equation}
where the 6-j symbols needed in Equation A1 may be found in Table 5 of Edmonds (1974).
The ratios must be summed over the degenerate transitions from the same
$J = 1 \, F_1^\prime \, F^\prime$ level (Table 5).

It is necessary to calculate the dependence of $R_i$ on the linewidth for closely spaced
hyperfine lines (see Turner 2001).  We have performed this calculation for the 
\nthp\ $J = 1 \rightarrow 0$ transitions assuming each hyperfine component is
Gaussian,
\begin{equation}
R_i(\Delta v) = \frac{ \sum_{j=1}^7  s_j \exp \left[ -4\ln 2 \frac{(v_j - v_i)^2}{\Delta v^2} \right] }
           { \sum_{j=1}^7  s_j \exp \left( -4\ln 2 \frac{v_j^2}{\Delta v^2} \right) } \;\; .
\end{equation}
The results are shown in Figure 10.
For sources with $\Delta v \leq 0.3$ km/s, the theoretical hyperfine ratios are
sufficient.  This is the case for L1498.  However, for \nthp\ observations of
sources with broader linewidths, the linewidth-corrected $R_i$ should be used.

\section*{Acknowledgments}

We wish to thank the staffs of the NRO, JCMT, CSO, and GBT for their excellent
assistance.  We are very grateful to Mario Tafalla and Derek Ward-Thompson
for providing us 1.2 mm and 170 \micron\ images of L1498, to
Luca Dore for providing us his calculation of the \nthp\ $J = 3 \rightarrow 2$
hyperfine frequencies, and to Antonio Crapsi for providing IRAM 30 m spectra of the 
\nthp\ $J = 3 \rightarrow 2$ transition.  We especially thank Ron Maddalena for assistance with
GBT calibration.  George Moellenbrock, Kumar Golap, and Joe McMullan provided excellent
assistance with Glish scripting in AIPS++.  Finally, we thank the referee for a
very detailed and helpful review.  NJE acknowledges support from Grant
AST-030725 from the National Science Foundation and NASA Grants NAG5-10488 and
NNG04GG24G.

%%%%%%%%%%%%%%%%%%%%%% References %%%%%%%%%%%%%%%%%%%%%%%%%%%%%%%%%%

%comment next line to include figures
%\end{document}
%%%%%%%%%%%%%%%%%% Figure captions %%%%%%%%%%%%%%%%%%%%%%%%%%%%%%%%%%

%%%%%%%%%%%%%%%%%% Figure 1 %%%%%%%%%%%%%%%%%%%%%%%%%%%%%%%%%%%%%%%%%%

\begin{figure}
\figurenum{1}
\epsscale{0.75}
\plotone{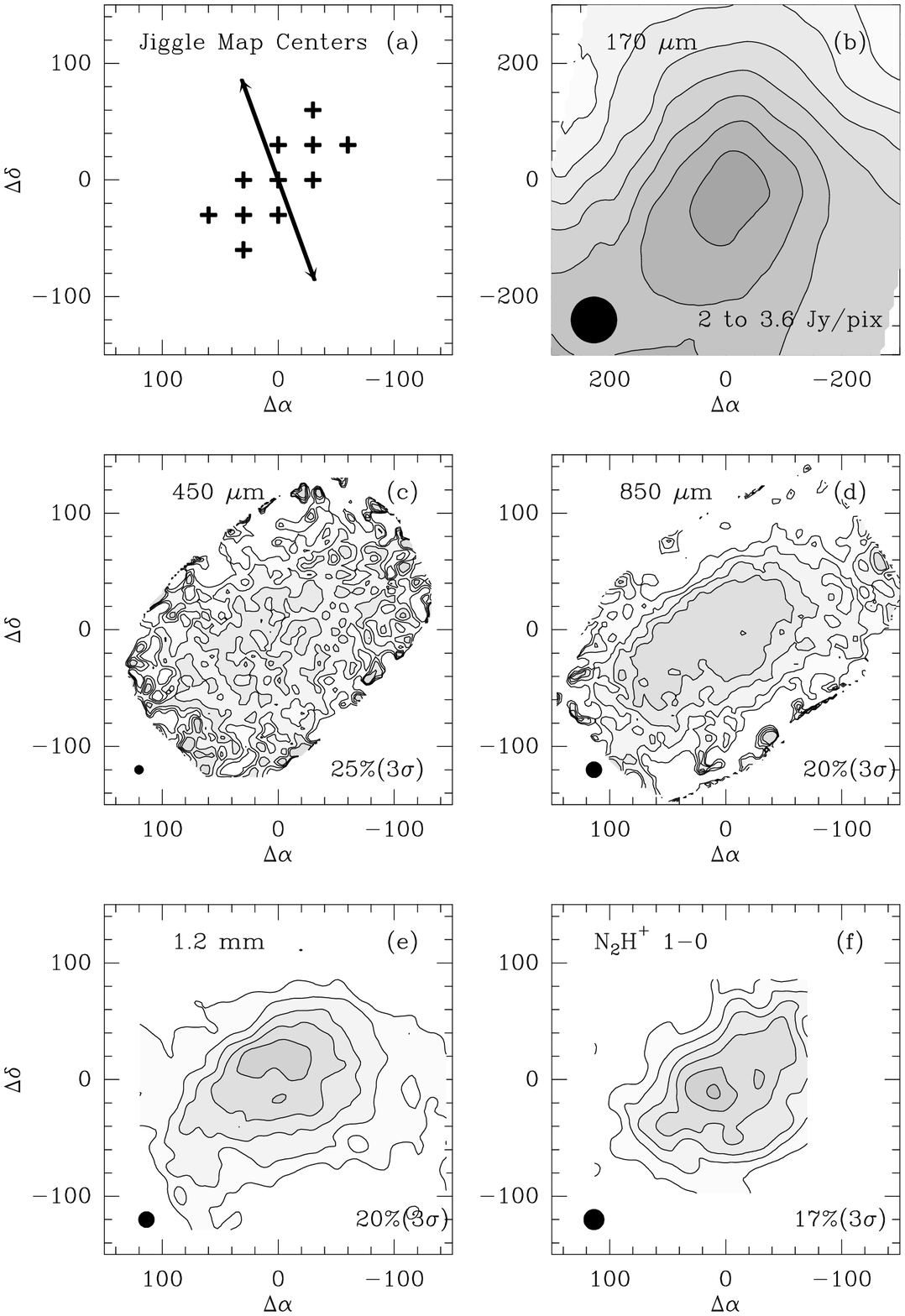}
\figcaption{ 
The jiggle map pointing centers are plotted with the chop direction (a).
The (0,0) position is J2000.0 $\alpha =$ $4^h$ $10^m$ $52.5^s$,
$\delta =$ $+25$\degree\ $9$\arcmin\ $55$\arcsec .
Contour maps of L1498 at 170 \micron\ (b; Ward-Thompson et al. 2002), 450 \micron\ (c), 
850 \micron\ (d), 
1200 \micron\ (e; Tafalla et al. 2004), and $I$(\ta ) 
of \nthp\ $J=1\rightarrow 0$ (f).  
The contour levels are as follows 
(lowest contour and contour increment
in percentage of the peak flux):
(\nthp ) 17\%(3$\sigma$),
(1200\micron ) 20\%(3$\sigma$) increasing by 20\% ;
(850\micron )  20\%(3$\sigma$) increasing by 20\% ;
(450\micron )  25\%(3$\sigma$) increasing by 25\% ; and
(170\micron )  2.0 to 3.6 Jy/pixel increasing by 0.2 Jy/pixel.
\textbf{The 170 \micron\ image is plotted on a scale
that is twice as large as the other images}.
%Contours near the edge of the maps should be ignored due to 
%noisy pixels, less integration time, and inability of the plotting
%package to handle irregular edges.
}
\end{figure}

%%%%%%%%%%%%%%%%%% Figure 2 %%%%%%%%%%%%%%%%%%%%%%%%%%%%%%%%%%%%%%%%%%

\begin{figure}
\figurenum{2}
\plotone{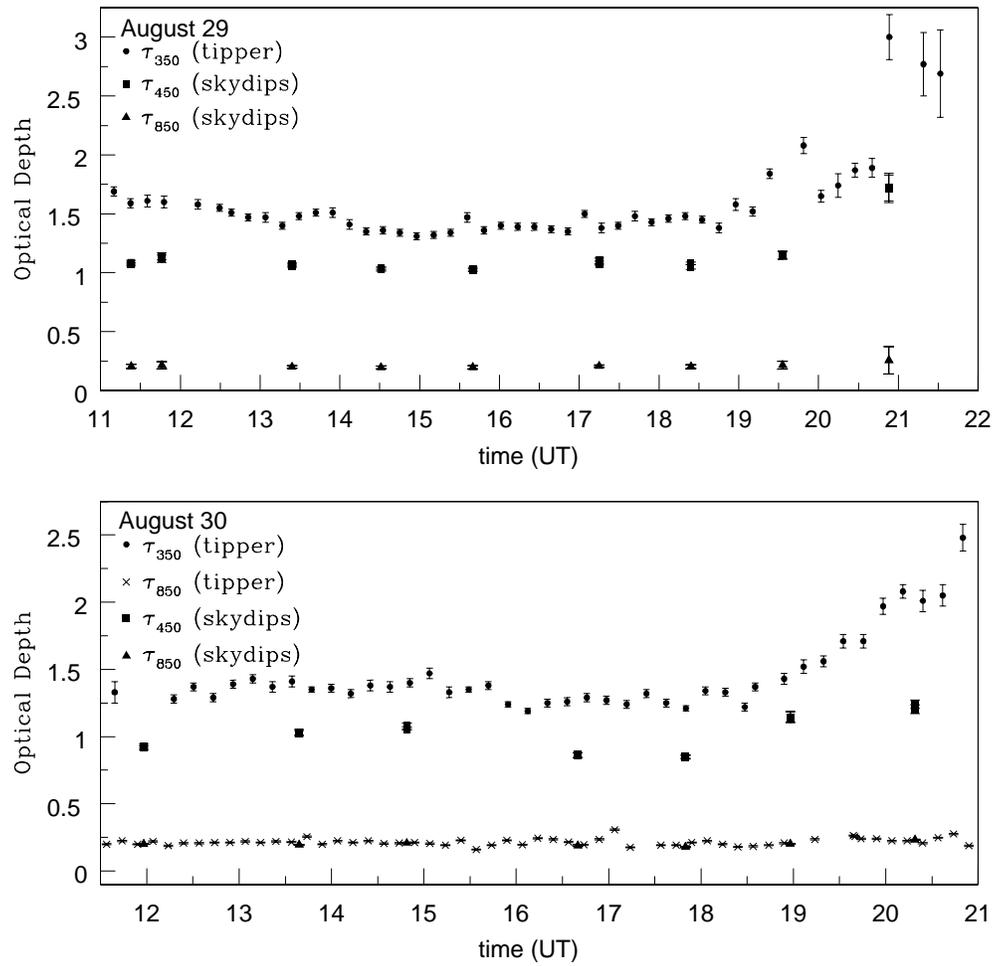}
\figcaption{ Opacity at 350, 450 and 850 \micron\ for August 29 and 30, 1998.  The
opacity was very stable throughout both nights except for 450 \micron\ 
observations during the final two hours of the morning shift (after sunrise).}
\end{figure}

%\begin{figure}
%\figurenum{4}
%\plotone{radprof.eps}
%\figcaption{ Radial profiles of L1498 at 850 and 450 \micron .
%The centroid is (11.1\arcsec ,-15.8\arcsec ).  The beam profile
%determined from jiggle maps of Uranus is shown as a dashed line.
%The solid red lines in the 850 \micron\ plot indicate the sector radial
%profiles along the major and minor axes.}
%\end{figure}

%%%%%%%%%%%%%%%%%% Figure 3 %%%%%%%%%%%%%%%%%%%%%%%%%%%%%%%%%%%%%%%%%%

\begin{figure}
\figurenum{3}
\plotone{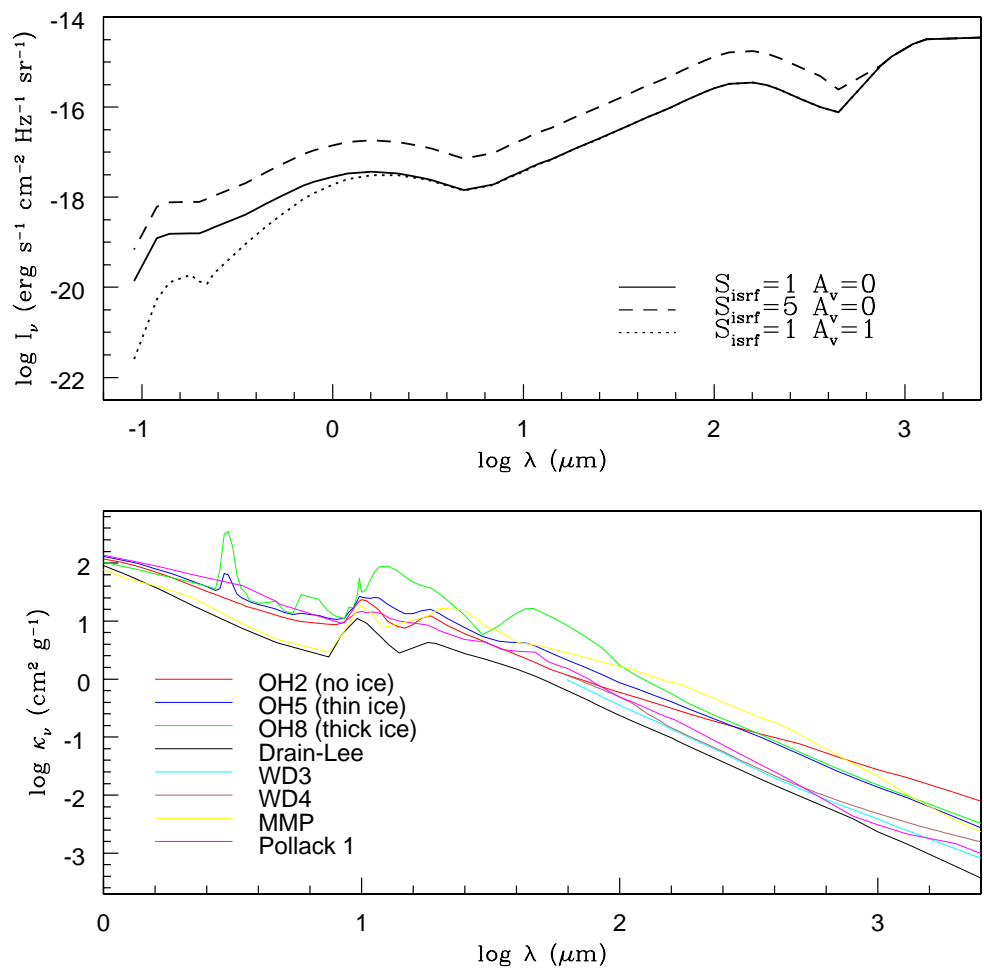}
\figcaption{The specific intensity of the ISRF with variations in $s_{isrf}$ and A$_{\rm{v}}$
(top panel).  The dust mass opacities (total gas + dust mass) used in dust models (bottom
panel).  A $M_{\rm{gas}}/M_{\rm{dust}} = 100$ was assumed.}
\end{figure}

%%%%%%%%%%%%%%%%%% Figure 4 %%%%%%%%%%%%%%%%%%%%%%%%%%%%%%%%%%%%%%%%%%

\begin{figure}
\figurenum{4}
\epsscale{0.75}
\plotone{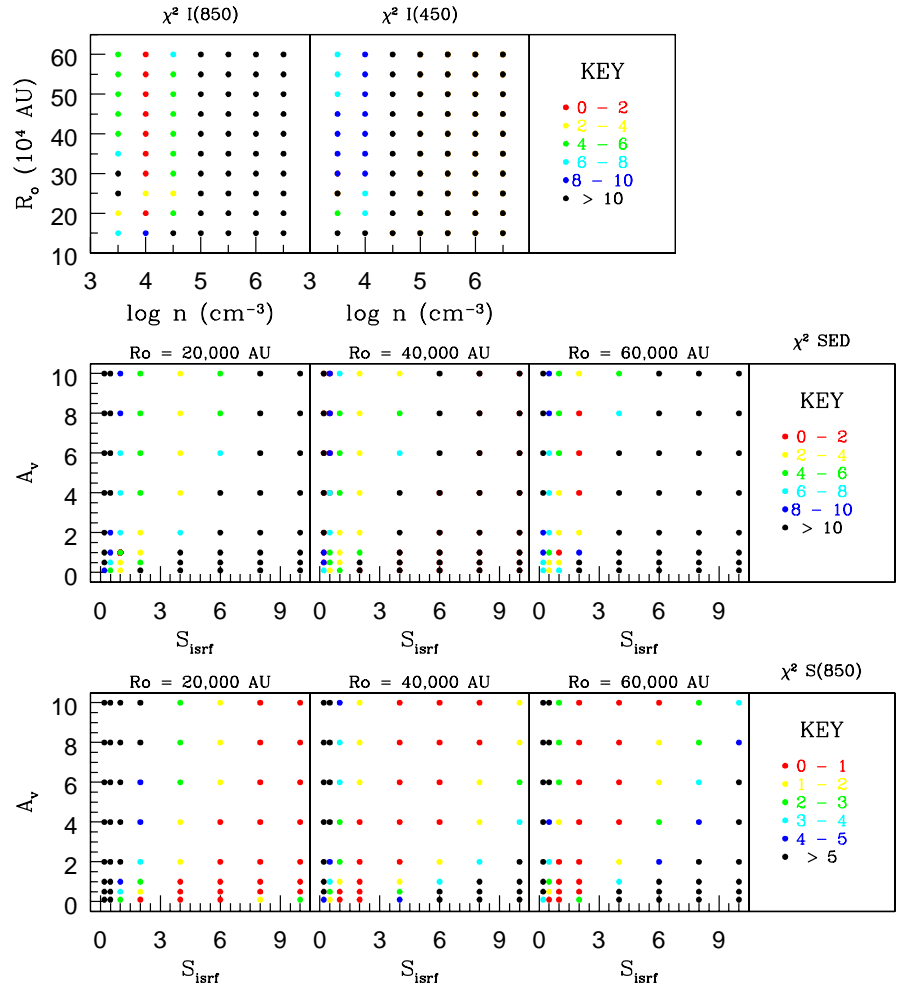}
\figcaption{ Color coded $\chi^2_r$ are plotted for the radiative transfer model grid for 
OH5 dust.  $\chi^2 (I_{850})$ and $\chi^2 (I_{450})$ are plotted for a grid of $n_c$ and $R_o$
(top row).  $\chi^2 (SED)$ (middle row) and $\chi^2 (S_{850})$ (bottom row) are plotted 
for grids of \sisrf\ and \av\ with outer radii
of 20,000, 40,000, and 60,000 AU.  Note that red points represent the
best $\chi^2_r$. }
\end{figure}

%%%%%%%%%%%%%%%%%% Figure 5 %%%%%%%%%%%%%%%%%%%%%%%%%%%%%%%%%%%%%%%%%%

%\begin{figure}
%\figurenum{5}
%\epsscale{0.75}
%\plotone{gridc.eps}
%\figcaption{ Radiative transfer model grid for OH5 dust. 
%Color coded $\chi^2_r$ for the azimuthally averaged, radial intensity profiles
%are plotted on the top row.  Color coded $\chi^2_r$ for the SED and $850$ \micron\
%flux are plotted.  The best-fitted model is found my finding the intersection of
%sets with $\chi^2_r \leq 1$.}
%\end{figure}

%%%%%%%%%%%%%%%%%% Figure 6 %%%%%%%%%%%%%%%%%%%%%%%%%%%%%%%%%%%%%%%%%%

%\begin{figure}
%\figurenum{6}
%\epsscale{0.75}
%\plotone{gridcc.eps}
%\figcaption{ Color coded $\chi^2_r$ for grids of radiative transfer models.}
%\end{figure}

%%%%%%%%%%%%%%%%%% Figure 5%%%%%%%%%%%%%%%%%%%%%%%%%%%%%%%%%%%%%%%%%%

\begin{figure}
\figurenum{5}
\epsscale{0.75}
\plotone{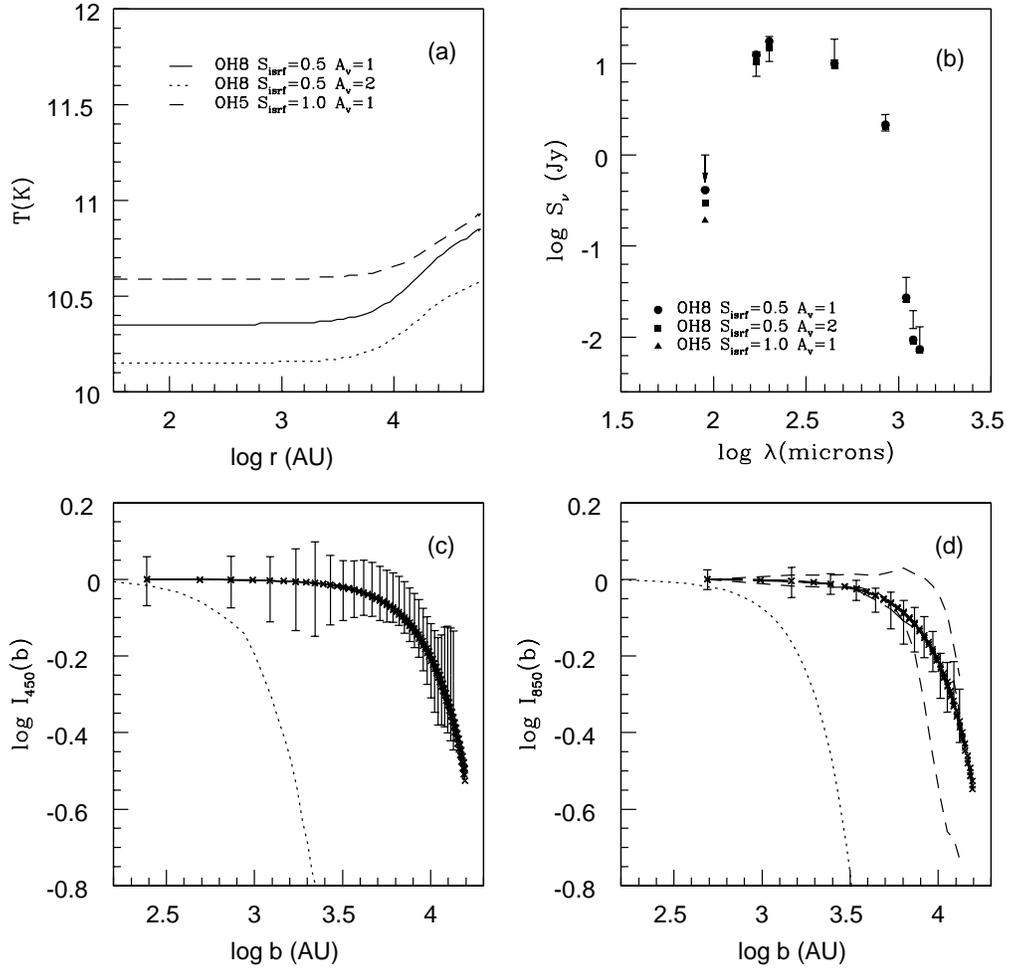}
\figcaption{Dust temperature profile (panel a), SED (b), $I_{850}(b)$ profile (c), and 
$I_{450}(b)$ profile (d) for the best fitted models.  The Uranus
beam profile is shown in the 850 and 450 \micron\ profile.  The sector-average
radial profiles at 850 \micron\ are plotted as dashed-lines (d).}
\end{figure}

%%%%%%%%%%%%%%%%%% Figure 6 %%%%%%%%%%%%%%%%%%%%%%%%%%%%%%%%%%%%%%%%%%

\begin{figure}
\figurenum{6}
\epsscale{0.75}
\plotone{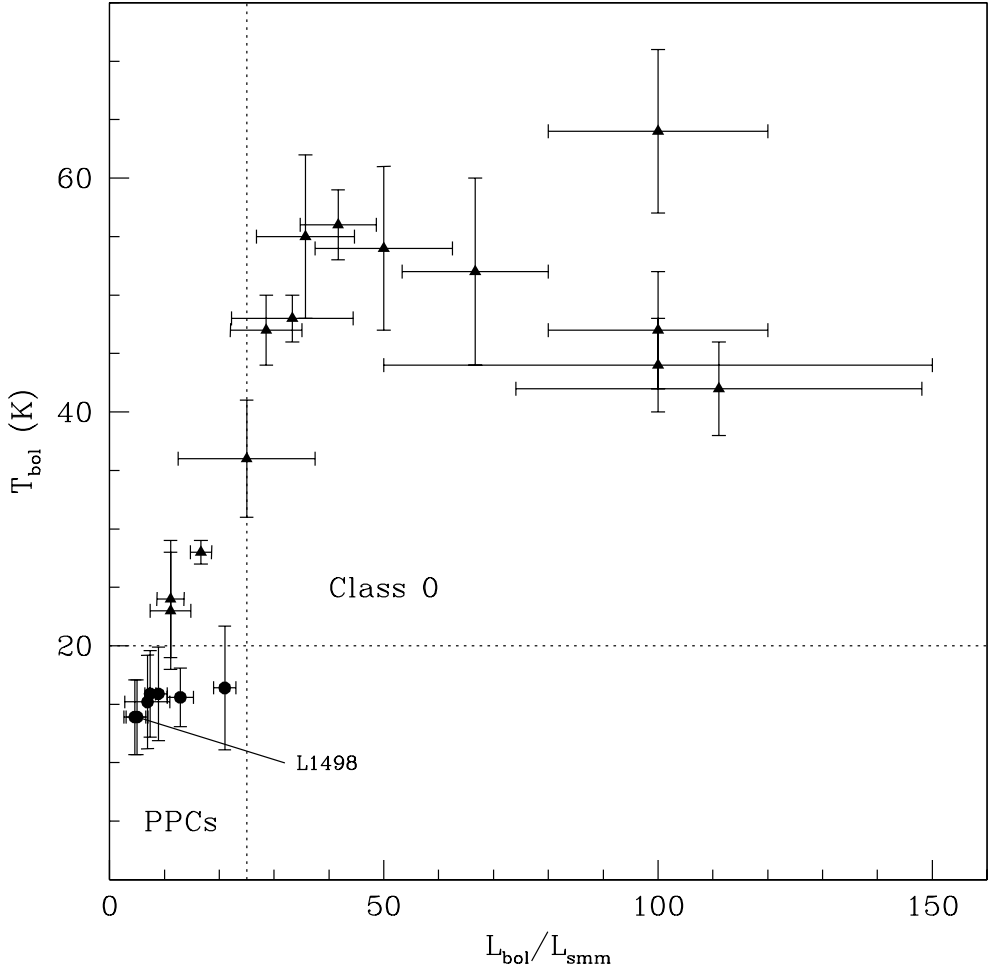}
\figcaption{The standard evolutionary indicators for low-mass protostars are
plotted: \tbol\ vs. \lbolsmm .  PPCs are plotted as circles and Class 0 sources
(Shirley et al. 2000, Young et al. 2003) are plotted as triangles.
There is no correlation between these
indicators for PPCs; however, all observed PPCs have \tbol\ $< 20$ K and 
\lbolsmm\ $< 25$.
}
\end{figure}

%%%%%%%%%%%%%%%%%% Figure 7 %%%%%%%%%%%%%%%%%%%%%%%%%%%%%%%%%%%%%%%%%%

\begin{figure}
\figurenum{7}
\epsscale{0.75}
\plotone{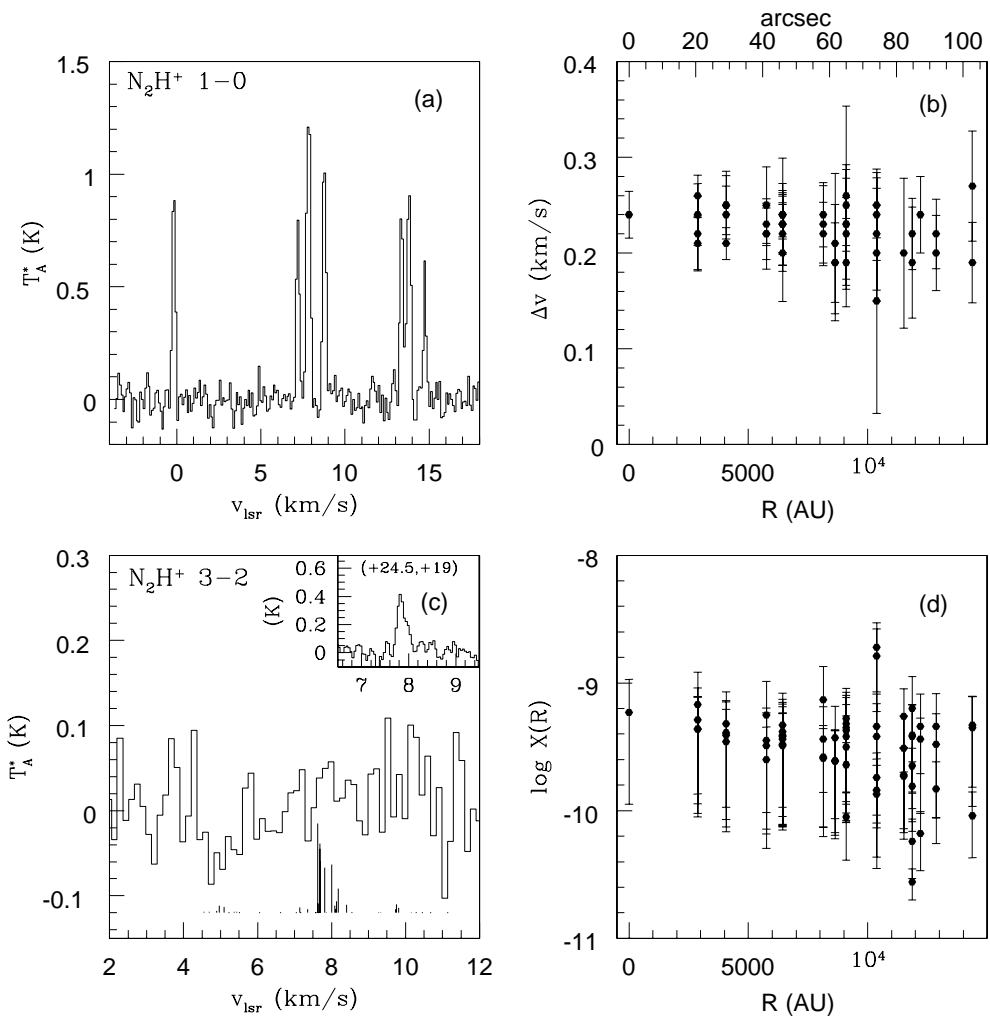}
\figcaption{The \nthp\ $J=1 \rightarrow 0$ spectrum (a) towards the SCUBA continuum
centroid (J2000.0 $\alpha =$ $4^h$ $10^m$ $53.3^s$,
$\delta =$ $+25$\degree\ $9$\arcmin\ $39$\arcsec ).  
The linewidth of the isolated hyperfine component ($J F_1 F = 
1 0 1 \rightarrow 0 1 2$) is plotted versus radius (b).  The 
\nthp\ $J=3 \rightarrow 2$ spectrum at the SCUBA continuum centroid
is shown in panel (c) with a theoretical
hyperfine stick spectrum.  The \nthp\ $J=3 \rightarrow 2$ detection 
(Crapsi et al. 2004) toward the
1.2 mm continuum peak (J2000.0 $\alpha =$ $4^h$ $10^m$ $51.5^s$,
$\delta =$ $+25$\degree\ $9$\arcmin\ $58$\arcsec ) is shown in the inset (c).
The abundance of \nthp\ versus radius is plotted in panel (d) with
statistical errorbars ($\sigma_I$, $\sigma_{\tau}$, \& $\sigma_{S_{\nu}}$).  
No significant variations
of the linewidth and abundance are observed on scales out to 15000 AU.
}
\end{figure}

%%%%%%%%%%%%%%%%%% Figure 8 %%%%%%%%%%%%%%%%%%%%%%%%%%%%%%%%%%%%%%%%%%

\begin{figure}
\figurenum{8}
%\epsscale{0.8}
%\plotone{evolseq.eps}
\centering
\vspace*{7.8cm}
\includegraphics{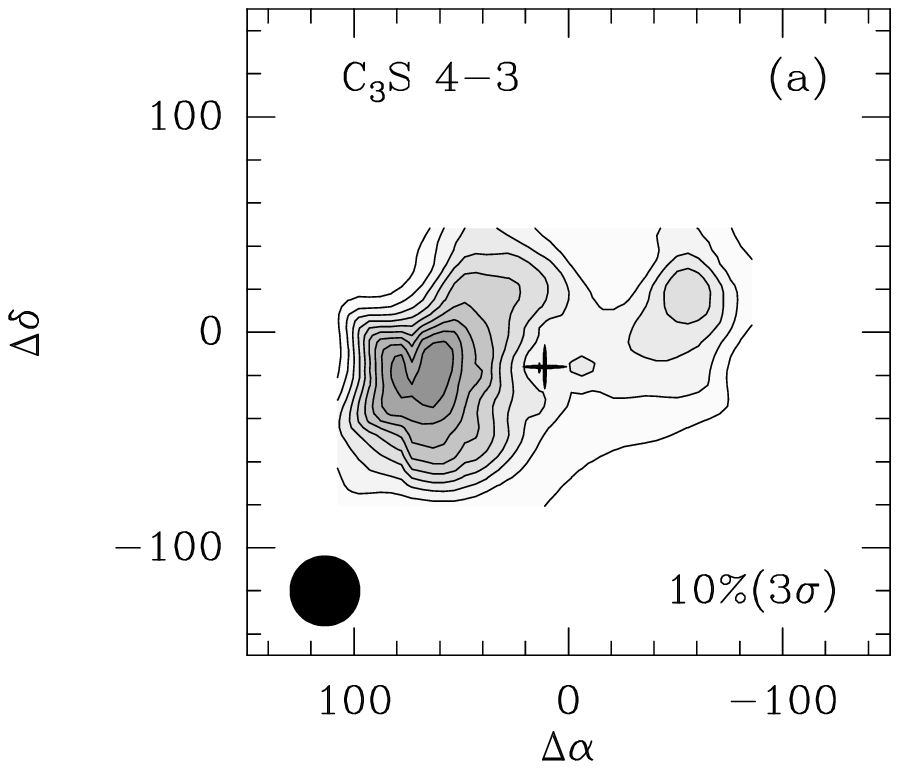}
\includegraphics{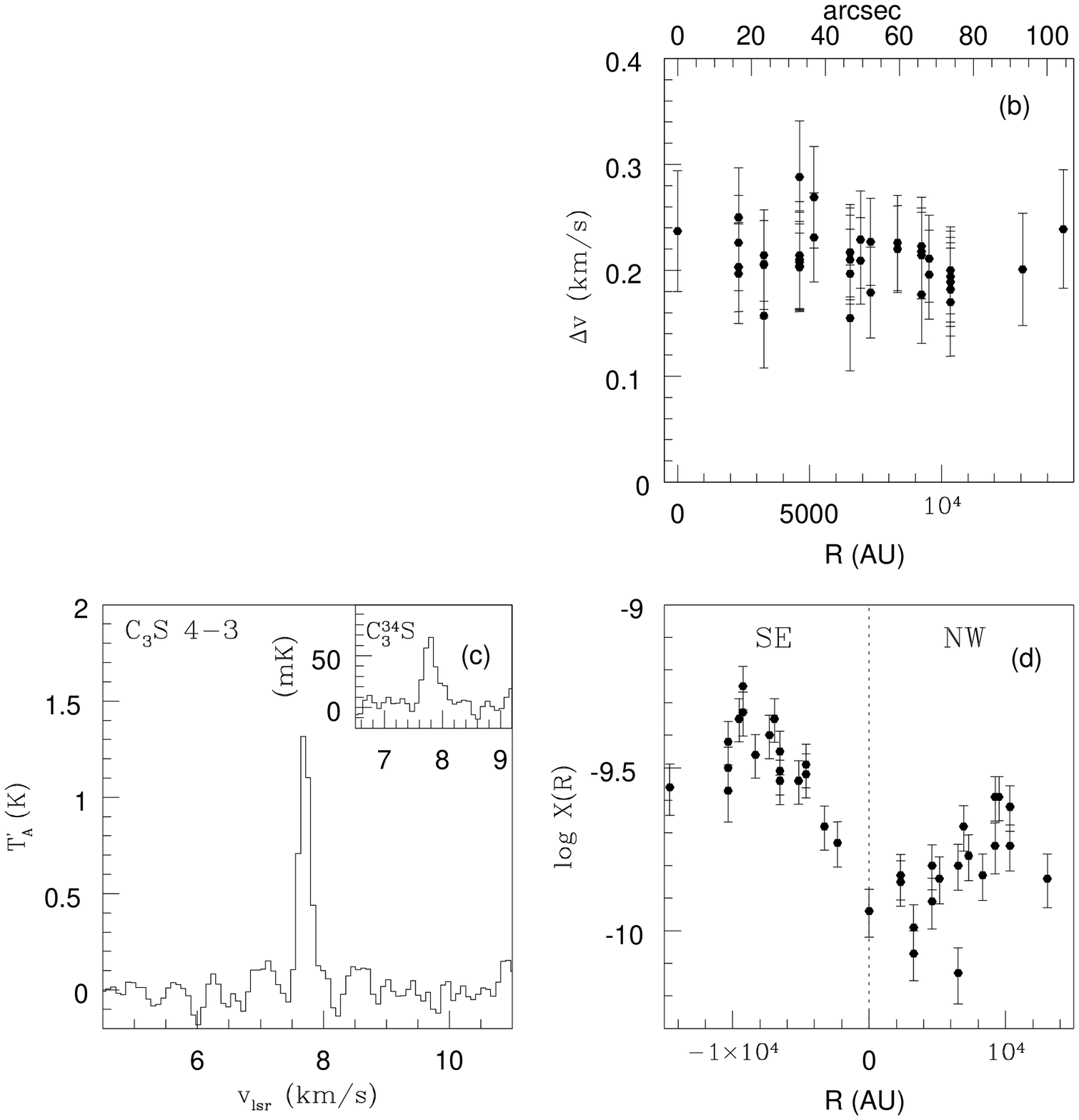}
\vskip 3.25in
\figcaption{The \cts\ $J = 4 \rightarrow 3$ integrated intensity map is plotted in panel (a)
with a cross at the position of the SCUBA dust continuum centroid. 
The (0,0) position is identical to Figure 1 (J2000.0 $\alpha =$ $4^h$ $10^m$ $52.5^s$,
$\delta =$ $+25$\degree\ $9$\arcmin\ $55$\arcsec ). 
The variation of linewidth is shown with radius (b).
The \cts\ and C$_3$$^{34}$S (inset) $J=4 \rightarrow 3$ spectra (c) towards the SE \cts\ peak
(J2000.0 $\alpha =$ $4^h$ $10^m$ $56.9^s$, $\delta =$ $+25$\degree\ $9$\arcmin\ $22\farcs 5$ ). 
The abundance of \cts\ versus radius in the major axis sector (\S2.1.2) is plotted in panel (d)
with statistical errorbars ($\sigma_I$ \& $\sigma_{S_{\nu}}$).  $R = 0$ is
the dust continuum centroid position; $-R$
refers to positions to the SE.  
Significant depletion is seen toward the dust continuum centroid.
}
\end{figure}

%%%%%%%%%%%%%%%%%% Figure 9 %%%%%%%%%%%%%%%%%%%%%%%%%%%%%%%%%%%%%%%%%%

\begin{figure}
\figurenum{9}
\epsscale{0.8}
\plotone{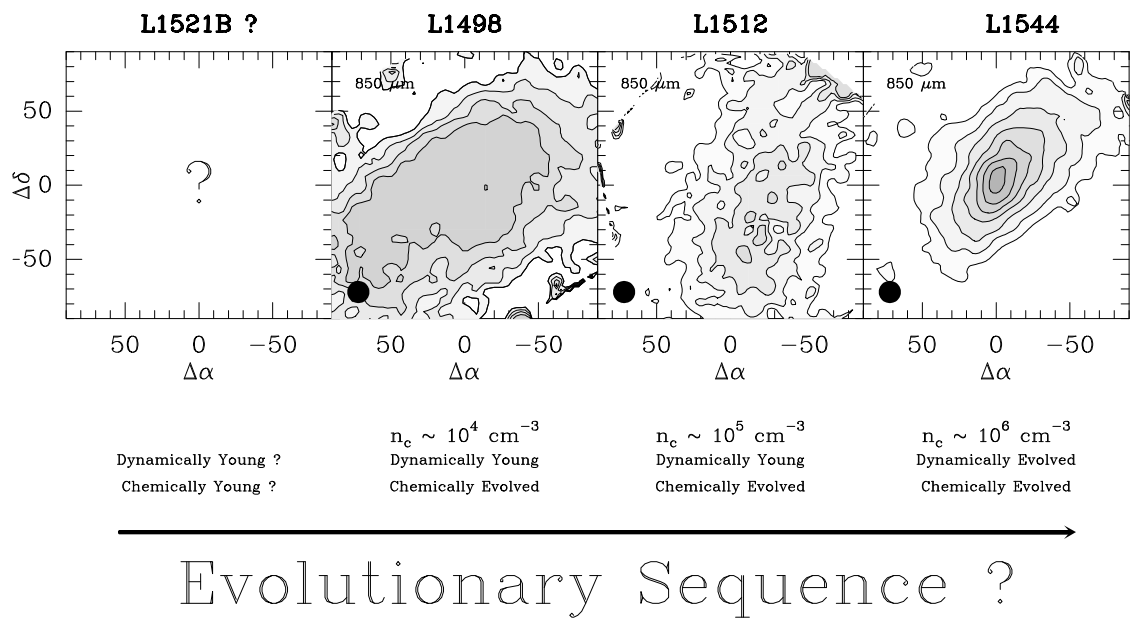}
\figcaption{Is this a possible evolutionary sequence for PPCs in Taurus-Auriga?  The 850 \micron\
images of L1512 and L1544 are from Shirley et al. (2000) while the modeled central
densities are from Evans et al. (2001).  Lee et al. (2003) have characterized the
chemical states of L1512 and L1544.  There are no published or archive submillimeter detections 
of the nascent pre-protostellar candidate, L1521B.
}
\end{figure}

%%%%%%%%%%%%%%%%%% Figure 10 %%%%%%%%%%%%%%%%%%%%%%%%%%%%%%%%%%%%%%%%%%

\begin{figure}
\figurenum{10}
\epsscale{0.75}
\plotone{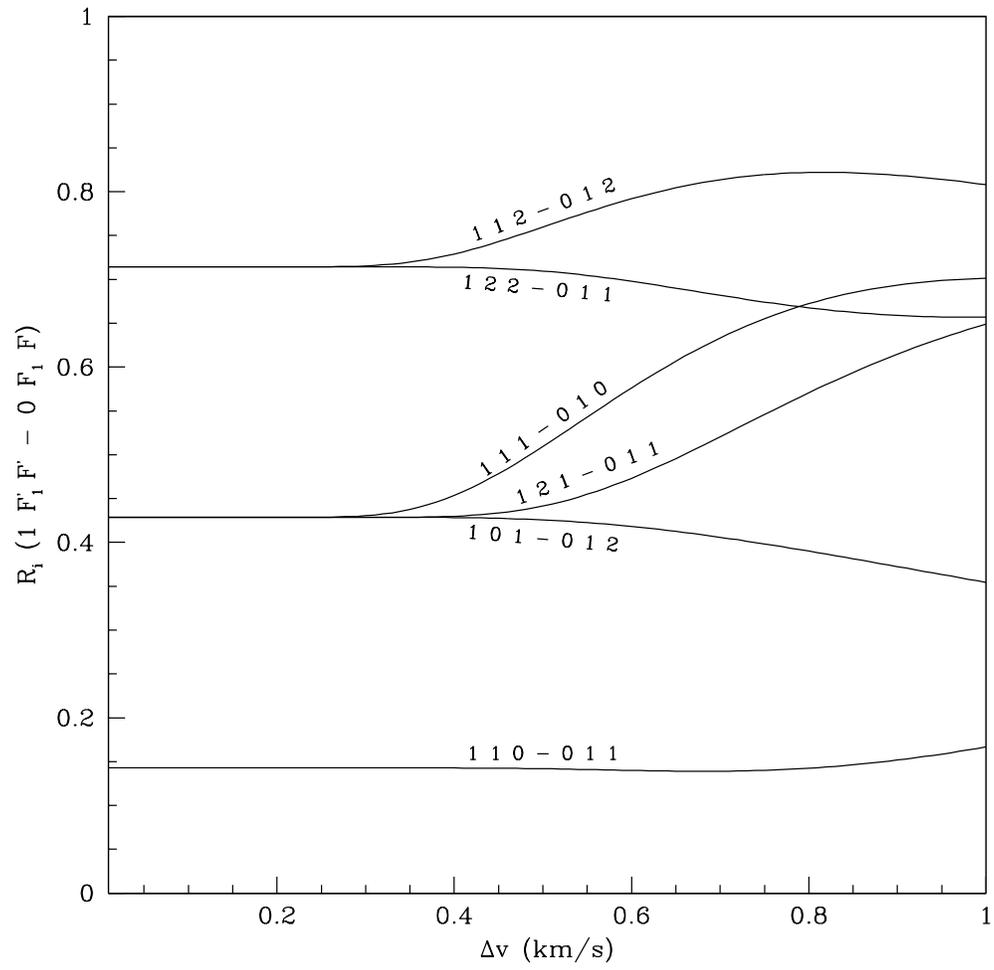}
\figcaption{The variation of the ratio of relative strengths for \nthp\
J = $1 \rightarrow 0$ hyperfine lines with linewidth.  The theoretical ratios are
valid for $\Delta v \leq 0.3$ km/s.
}
\end{figure}

%%%%%%%%%%%%%%%%%% Table 1 %%%%%%%%%%%%%%%%%%%%%%%%%
\begin{deluxetable}{lccl}
\tablecolumns{4}
\footnotesize
\tablecaption{L1498 Spectral Energy Distribution \label{tab1}}
\tablewidth{0pt} 
\tablehead{
\colhead{$\lambda$ (\micron)}   &
\colhead{$S_{\nu}$ (Jy)}        &
\colhead{$\theta$ (\arcsec)}    &
\colhead{Ref.}		                           
}
\startdata 
90   & $\leq$ 1.0           & 150       & 1 \\
170  & 10.4 (...)	    & 150 	& 1 \tablenotemark{a}\\
200  & 15.2 (...)  	    & 150 	& 1 \tablenotemark{a}\\
450  & 0.700 (0.080)	    & 18 	& 2 \\ 
450  & 2.69 (0.59)	    & 40 	& 3 \\ 
450  & 14.70 (2.42)	    & 120 	& 3 \tablenotemark{a}\\ 
800  & 0.120 (0.018)	    & 18 	& 2 \\ 
850  & 0.376 (0.021)	    & 40 	& 3 \\ 
850  & 2.30 (0.08)	    & 120 	& 3 \tablenotemark{a}\\ 
1100 & 0.035 (0.006)	    & 18 	& 2 \tablenotemark{a}\\
1200 & 0.016 (0.004)	    & 12	& 4 \tablenotemark{a}\\ 
1300 & 0.010 (0.003)	    & 12 	& 2 \tablenotemark{a}\\
\enddata
\tablenotetext{a}{Fluxes used in calculation of $\chi^2_{\rm{SED}}$.}
\tablerefs{1. Ward-Thompson et al. 2002, 2. Ward-Thompson et al. 1994, 3. This paper,
4. Tafalla et al. 2004}
\end{deluxetable}

%%%%%%%%%%%%%%%%%% Table 2 %%%%%%%%%%%%%%%%%%%%%%%%%
\begin{deluxetable}{llccc}
\footnotesize
\tablecolumns{5}
\tablecaption{Dust Opacity Properties \label{tab2}}
\tablewidth{0pt} 
\tablehead{
\colhead{Model}					&
\colhead{Description}           		&
\colhead{$\beta_{\rm{smm}}$\tablenotemark{a}}	&
\colhead{$\kappa_{\nu}(850)$}        		&
\colhead{Ref.}       				\\
\colhead{Name}					&
\colhead{}					&
\colhead{}					&
\colhead{(cm$^2$ g$^{-1}$)}			& 
\colhead{}          
}
\startdata 
Draine-Lee	& Silicates + Carbonaceous grains
	& 2.0	& 0.003	        & 1 \\
WD3             & Silicate + Carbonaceous + PAH ISM fit
        & 1.8  & 0.005         & 2 \\
WD4             & Silicate + Carbonaceous + PAH ISM fit  
        & 1.6  & 0.006           & 2 \\
Pollack1        & Silicates + Organic C  
        & 2.2  & 0.004         & 3 \\ 
MMP		& Empirical Fit to ISM
	& 2.3	& 0.030		& 4 \\
OH2		& Coagulated for $10^5$ yrs. no ice mantles
	& 1.3	& 0.034		& 5 \\
OH5		& Coagulated for $10^5$ yrs. thin ice mantles 
	& 1.8	& 0.018 	& 5 \\
OH8		& Coagulated for $10^5$ yrs. thick ice mantles
	& 1.9	& 0.022		& 5 \\
\enddata

\tablenotetext{a}{$\kappa \sim \nu^{\beta}$ where $\beta_{\rm{smm}}$ is
determined by a linear regression over $\lambda \in [350 \micron , 1300 \micron ]$.}
\tablerefs{1. Draine \& Lee 1984, 2. Weingartner \& Draine 2001, 3. Pollack et al. 1994, 
4. Mathis, Metzger, \& Panagia 1983, 5. Ossenkopf \& Henning 1994} 
\end{deluxetable}

%%%%%%%%%%%%%%%%%% Table 3 %%%%%%%%%%%%%%%%%%%%%%%%%
\begin{deluxetable}{lcccccccc}
\tablecolumns{9}
\footnotesize
\tablecaption{Best-fitted Models\tablenotemark{a}\label{tab3}}
\tablewidth{0pt} 
\tablehead{
\colhead{$\log n_c$}    &
\colhead{$R_o$}    &
\colhead{\sisrf\ } &
\colhead{\av\ }    &
\colhead{$\kappa_{\nu}$} &
\colhead{$\chi^2_r(I_{850})$} &
\colhead{$\chi^2(S_{850})$} & 
\colhead{$\chi^2_r(SED)$}  \\
\colhead{(\cmv ) }    &
\colhead{($10^4$ AU)}    &
\colhead{} &
\colhead{(mag)}    &
\colhead{} &
\colhead{} &
\colhead{} & 
\colhead{} 		                           
}
\startdata 
4.0  & 60 & 0.5 & 1.0\tablenotemark{b} & OH8 & 0.67 & 0.12 & 0.84 \\
4.0  & 60 & 2.0 & 4.0 & OH5 & 0.52 & 0.10 & 0.90 \\
4.0  & 60 & 2.0 & 6.0 & OH5 & 0.52 & 0.26 & 0.84 \\
4.0  & 60 & 0.5 & 2.0\tablenotemark{b} & OH8 & 0.66 & 0.28 & 0.88 \\
4.0  & 60 & 1.0 & 1.0\tablenotemark{b} & OH5 & 0.58 & 0.35 & 0.90 \\
4.0  & 60 & 2.0 & 8.0 & OH5 & 0.52 & 0.46 & 0.97 \\
%\hline
%4.0  & 60 & 0.5 & 1.0 & OH5 & 0.82 & 0.52 & 2.01 \\
%4.0  & 40 & 2.0 & 4.0 & OH5 & 0.67 & 0.76 & 2.07 \\
%4.0 & 40 & 1.0 & 10.0& OH8 & 0.84 & 0.43 & 2.10 \\
%4.0 & 40 & 0.5 & 0.5 & OH8 & 0.94 & 0.46 & 2.11 \\
%4.0 & 40 & 0.5 & 1.0 & OH8 & 0.88 & 0.76 & 2.11 \\
%4.0 & 60 & 0.5 & 0.5 & OH8 & 0.70 & 0.02 & 2.12 \\
%4.0 & 60 & 0.5 & 4.0 & OH8 & 0.66 & 0.52 & 2.12 \\
%4.0 & 40 & 1.0 & 0.5 & OH5 & 0.91 & 0.55 & 2.17 \\
%4.0 & 40 & 1.0 & 8.0 & OH8 & 0.84 & 0.31 & 2.32 \\
%4.0 & 40 & 2.0 & 2.0 & OH5 & 0.69 & 0.28 & 2.59 \\
%4.0 & 40 & 1.0 & 6.0 & OH8 & 0.84 & 0.18 & 2.78 \\
\enddata
\tablenotetext{a}{Models for which $\left\{ \chi^2_r(I_{850}) \in [0,1] \right\} \; \bigcap \; 
\left\{\chi^2_r(S_{850}) \in [0,1] \right\} \; \bigcap \;
\left\{\chi^2_r(SED) \in [0,1] \right\}$.} 
%\tablenotetext{a}{Models for which $\left\{ \chi^2(I_{850}), \chi^2(S_{850}), \chi^2(SED) \right\}
%\; \in [0,1]$.} 
\tablenotetext{b}{Model is consistent with Cambr\'{e}sy 1999 optical extinction constraint that
\av\ $\leq 3$ mag.}
\end{deluxetable}

%%%%%%%%%%%%%%%%%% Table 4 %%%%%%%%%%%%%%%%%%%%%%%%%
%\begin{deluxetable}{lccc}
%\footnotesize
%\tablecolumns{5}
%\tablecaption{Characteristic Temperatures \label{tab4}}
%\tablewidth{0pt} 
%\tablehead{
%\colhead{T}					&
%\colhead{Description}         		        &
%\colhead{Equation}				&
%\colhead{Value}        				\\
%\colhead{}					&
%\colhead{}					&
%\colhead{}					&
%\colhead{(K)}			
%}
%\startdata 
%\tbol	& Bolometric 	& $\frac{h\zeta(4)}{4k\zeta(5)}\frac{\int_{0}^{\infty}\nu S_{\nu} d\nu}{\int_{0}^{\infty}S_{\nu} d\nu}$ 
%       	& 13.9 (3.1) \\  \\
%$T_{gb}$  & Greybody  	& min$\left[\sum_{i=1}^{N}F_{i}(\nu ) - S_{i}(\nu )\right]$
%	& ...  \\ \\
%$T_{iso}$ & Isothermal  & $\frac{h\nu}{k}\left[ 1 + \ln\left(\frac{2h\nu^3M_{env}\kappa_{\nu}}{S_{\nu}c^2D^2}\right) \right]$ 
%	& ...  \\ \\
%$T_{mass}$ & Mass-weighted & $\frac{\int_{0}^{\infty} n(r) T(r) r^2 dr}{\int_{0}^{\infty} n(r) r^2 dr}$
%	& ... \\ \\
%$T_{harm}$ & Harmonic  	& $\frac{\int_{0}^{\infty} n(r) dr}{\int_{0}^{\infty} \frac{n(r)}{T(r)} dr}$
%	& ... \\
%\enddata
%
%\end{deluxetable}

%%%%%%%%%%%%%%%%%% Table 4 %%%%%%%%%%%%%%%%%%%%%%%%%
\begin{deluxetable}{lccccccccccc}
\rotate 90
\tablecolumns{12}
\tablecaption{Properties of Nearby PPCs Observed with ISOPHOT\tablenotemark{a} and SCUBA \label{tab4}}
\tablewidth{0pt} 
\tablehead{
\colhead{Source}    &
\colhead{$\alpha_{\rm{J2000.0}}$ \tablenotemark{b}}    &
\colhead{$\delta_{\rm{J2000.0}}$ \tablenotemark{b}}    &
\colhead{Ref. \tablenotemark{c}} &
\colhead{D} &
\colhead{$2a \times 2b$} &
\colhead{$R_{1/2}$} &
\colhead{a/b}  & 
\colhead{\lbol }    &
\colhead{\lbolsmm } & 
\colhead{\tbol \tablenotemark{d}} &
\colhead{$M_{D}(10.5 \rm{K})$\tablenotemark{e}}  \\
\colhead{}    &
\colhead{($^h$~~$^m$~~$^s$~)~} &  
\colhead{($\degree$ ~\arcmin\ ~\arcsec)}   &
\colhead{} &
\colhead{(pc)} &
\colhead{(AU $\times$ AU)} & 
\colhead{(AU)} & 
\colhead{} &
\colhead{(\lsun )}    &
\colhead{} & 
\colhead{(K)} &
\colhead{(\msun )} 	                           
}
\startdata 
 L1498   & 04 10 53.3    &$+$25 09 39 & 1 & 140 & 27600 $\times$ 15000 & 10170 & 1.8 
              & 0.12 (0.02) & 5 (2)   & 13.9 (3.2) & 0.67 \\
 L1517B  & 04 55 18.1    &$+$30 37 48 & 3 & 140 & 11600 $\times$ 9700  & 5300 & 1.2  
              & 0.06 (0.01) & 18 (9)  & 15.2 (3.5) & 0.67 \\
 L1512  & 05 04 08.2    &$+$32 43 20  & 2 & 140 & 19000 $\times$ 8500  & 6350 & 2.2  
              & 0.10 (0.02) & 7 (4)   & 15.2 (4.0) & 0.53 \\
 L1544  & 05 04 17.1    &$+$25 10 48  & 2 & 140 & 10900 $\times$ 5700  & 3940  & 1.9  
              & 0.15 (0.02) & 5 (2)   & 13.9 (3.2) & 1.06 \\
 L183   & 15 54 08.5    &$-$02 52 32  & 3 & 150 & 11400 $\times$ 6200  & 4200  & 1.8  
              & 0.09 (0.01) & 13 (3)  & 15.6 (2.5) & 0.89 \\
 L1709A & 16 30 50.4    &$-$23 42 05  & 3 & 125 & 15800 $\times$ 10500 & 6440 & 1.5  
              & 0.09 (0.02) & ...     & 15.4 (4.9) & 0.98 \\
 L1689B & 16 34 49.2    &$-$24 38 07  & 2 & 125 & 9600  $\times$ 7700  & 4300  & 1.3  
              & 0.13 (0.02) & 7 (1)   & 15.9 (3.7) & 0.74 \\
 L63    & 16 50 14.4    &$-$18 06 17  & 3 & 160 & 12400 $\times$ 7800  & 4920  & 1.6  
              & 0.20 (0.04) & 21 (2)  & 16.4 (5.3) & 1.15 \\
 B68    & 17 22 39.0    &$-$23 49 57  & 4 & 95  & 9500 $\times$ 7500   & 5530 & 1.3  
              & 0.05 (0.02) & 5 (2)   & 14.5 (3.5) & 0.50 \\
 B133   & 19 06 08.0    &$-$06 52 52  & 2 & 200 & 14900 $\times$ 12100 & 6710 & 1.2  
              & 0.37 (0.06) & 9 (2)   & 15.9 (4.0) & 1.22 \\
\hline
$\mean{x}$ &            &             &   & 142 &                      & 5790 & 1.6 
              & 0.14        & 9.9     & 15.2       & 0.8 \\
$\mu_{1/2}(x)$\tablenotemark{f}&&     &   & 140 &                      & 5420 & 1.55  
              & 0.11        & 7.4     & 15.3       & 0.9 \\
$\sigma_x$&             &             &   & 27  &                      & 1830 & 0.4  
              & 0.09        & 6.0     & 0.9        & 0.3 \\ 
\enddata
\tablenotetext{a}{ISOPHOT fluxes from Ward-Thompson et al. 2002.}
\tablenotetext{b}{SCUBA 850 \micron\ continuum  FWHM contour centroid in J2000.0
coordinates.}
\tablenotetext{c}{Reference for SCUBA observations.}
\tablenotetext{d}{\tbol\ $= \frac{h\zeta(4)}{4k\zeta(5)}
\frac{\int_{0}^{\infty}\nu S_{\nu} d\nu}{\int_{0}^{\infty}S_{\nu} d\nu}$.}
%\tablenotetext{d}{$T_{iso}$ $= \frac{h\nu}{k}\left[ 1 + 
%\ln\left(\frac{2h\nu^3M_{env}\kappa_{\nu}}{S_{\nu}c^2D^2}\right) \right]$, where $S_{\nu}$ 
%and $M_{env}$ are determined for a $120$\arcsec\ aperture.}
\tablenotetext{e}{Dust-determined mass in a 120\as\ aperture.}
\tablenotetext{f}{Median of the sample.}
\tablerefs{1. This paper; 2. Shirley et al. 2000; 3. CADC SCUBA Archive; 4. Bianchi et al. 2003, Lai et al. 2003} 
\end{deluxetable}

%%%%%%%%%%%%%%%%%% Table A1 %%%%%%%%%%%%%%%%%%%%%%%%%
\begin{deluxetable}{lclccc}
\tablecolumns{6}
\footnotesize
\tablecaption{\nthp\ $J = 1 \rightarrow 0$ Hyperfine Transitions \label{tab5}}
\tablewidth{0pt} 
\tablehead{
\colhead{$J^{\prime}F_1^{\prime}F^{\prime} \rightarrow JF_1F$}    &
\colhead{$s_i$} &
\colhead{$J^{\prime}F_1^{\prime}F^{\prime} \rightarrow JF_1F$}    &
\colhead{$\nu$}                      &
\colhead{$\Delta v$}                 &
\colhead{$R_i$}                      \\
\colhead{}                 &
\colhead{}                 &
\colhead{Standard Notation}                 &
\colhead{(GHz)}                  &
\colhead{(km/s)}                 &
\colhead{}
}
\startdata
1 1 0 $\rightarrow$ 0 1 1   & $\frac{1}{27}$ & 1 1 0 $\rightarrow$ 0 1 1 &93.1716200 &$+$6.944 & $\frac{1}{7}$ \\
\hline
1 1 2 $\rightarrow$ 0 1 2   & $\frac{5}{36}$ & 1 1 2 $\rightarrow$ 0 1 2 & 93.1719168 & $+$5.988 &$\frac{5}{7}$\\
1 1 2 $\rightarrow$ 0 1 1   & $\frac{5}{108}$& \\
\hline
1 1 1 $\rightarrow$ 0 1 2   & $\frac{5}{108}$& \\
1 1 1 $\rightarrow$ 0 1 1   & $\frac{3}{108}$& \\
1 1 1 $\rightarrow$ 0 1 0   & $\frac{1}{27}$ & 1 1 1 $\rightarrow$ 0 1 0 & 93.1720533 &$+$5.549 &$\frac{3}{7}$ \\
\hline
1 2 2 $\rightarrow$ 0 1 2   & $\frac{5}{108}$& \\
1 2 2 $\rightarrow$ 0 1 1   & $\frac{5}{36}$ & 1 2 2 $\rightarrow$ 0 1 1 & 93.1734796 &$+$0.956 &$\frac{5}{7}$ \\
\hline
1 2 3 $\rightarrow$ 0 1 2   & $\frac{7}{27}$ & 1 2 3 $\rightarrow$ 0 1 2 & 93.1737767 & $+$0.000 & $1$ \\
\hline
1 2 1 $\rightarrow$ 0 1 2   & $\frac{1}{324}$\\
1 2 1 $\rightarrow$ 0 1 1   & $\frac{5}{108}$& 1 2 1 $\rightarrow$ 0 1 1 & 93.1739666 &$-$0.611 &$\frac{3}{7}$ \\
1 2 1 $\rightarrow$ 0 1 0   & $\frac{5}{81}$ \\ 
\hline
1 0 1 $\rightarrow$ 0 1 2   & $\frac{5}{81}$ & 1 0 1 $\rightarrow$ 0 1 2 & 93.1762650 &$-$8.011 &$\frac{3}{7}$ \\
1 0 1 $\rightarrow$ 0 1 1   & $\frac{1}{27}$ & \\
1 0 1 $\rightarrow$ 0 1 0   & $\frac{1}{81}$ & \\
\enddata
\end{deluxetable}

\end{document}